\documentclass[traditabstract]{aa} % for the abstract without structuration 
\usepackage{longtable,lscape}
\usepackage{graphicx}
\usepackage{natbib}
\usepackage{txfonts}
\usepackage{multirow}
\begin{document}

   \title{Hot HCN around young massive stars at $0.1''$ resolution}

   \author{R. Rolffs
          \inst{1,2}
          \and
          P. Schilke\inst{2}
          \and
          F. Wyrowski\inst{1}
         \and 
          C. Dullemond\inst{3,4}
          \and
          K. M. Menten\inst{1}
          \and
          S. Thorwirth\inst{2}
	  \and
	  A. Belloche\inst{1}
          }

   \institute{Max-Planck-Institut f\"ur Radioastronomie,
              Auf dem H\"ugel 69, 53121 Bonn, Germany\\
              \email{rrolffs@mpifr.de}
         \and
             I. Physikalisches Institut, Universit\"at zu K\"oln,
  Z\"ulpicher Stra\ss e 77, 50937 K\"oln, Germany\\
             \email{schilke@ph1.uni-koeln.de}
 	\and
	     Institut f\"ur Theoretische Astrophysik,  Universit\"at Heidelberg, 
	Albert-Ueberle-Str. 2, 69120 Heidelberg, Germany
	\and
	     Max-Planck-Institut f\"ur Astronomie,
             K\"onigstuhl 17, 69117 Heidelberg, Germany
         }

   \date{Received 20 January 2011 / Accepted 3 March 2011}

  \abstract{
% Context
Massive stars form deeply embedded in dense molecular gas, which they stir and heat up and ionize. During an early phase, the ionization is confined to hypercompact H{\sc ii} regions, and the stellar radiation is entirely absorbed by dust, giving rise to a hot molecular core. 
% Aims
To investigate the innermost structure of such high-mass star-forming regions, 
% Methods
we observed vibrationally excited HCN (via the direct $\ell$-type transition of $v_2$=1, $\Delta J$=0,  $J$=13, which lies 1400 K above ground) toward the massive hot molecular cores G10.47+0.03, SgrB2-N, and SgrB2-M with the Very Large Array (VLA) at 7 mm, reaching a resolution of about 1000 AU ($0.1''$). 
% Results
We detect the line both in emission and in absorption against H{\sc ii} regions. The latter allows to derive lower limits on the column densities of hot HCN, which are several times $10^{19}$ cm$^{-2}$. We see indication of expansion motions in G10.47+0.03 and detect velocity components in SgrB2-M at 50, 60, and 70 km~s$^{-1}$ relative to the Local Standard of Rest. The emission originates in regions of less than 0.1 pc diameter around the hypercompact H{\sc ii} regions G10.47+0.03 B1 and SgrB2-N K2, and reaches brightness temperatures of  more than 200 K. Using the three-dimensional radiative transfer code RADMC-3D, we model the sources as dense dust cores heated by stars in the H{\sc ii} regions, and derive masses of hot ($>$300 K) molecular gas of more than 100 solar masses (for an HCN fractional abundance of 10$^{-5}$), challenging current simulations of massive star formation.
% Conclusions
Heating only by the stars in the H{\sc ii} regions is sufficient to produce such large quantities of hot molecular gas, provided that dust is optically thick to its own radiation, leading to high temperatures through diffusion of radiation. 
}

   \maketitle
%
%________________________________________________________________

\section{Introduction}

The formation of massive stars occurs deep inside dense molecular cloud cores. The high luminosity of massive (proto)stars stirs and heats up the surrounding gas,  leading to the evaporation of ice mantles around dust grains and to high degrees of excitation of the molecules. This stage, just prior to ionization and destruction of the molecular gas, is characterized by strong line and dust emission from a compact region, and is called a hot molecular core \citep[e.g.][]{Cesaroni05}. Since these sources are short-lived, they are rare and hence far away, which hampers the determination of their physical structure (density, temperature, velocity, molecular abundances). Nonetheless, such knowledge is a key information for understanding the process of massive star formation \citep{ZinneckerYorke} and for the interpretation of molecular line surveys \citep[e.g.][]{Schilke06,Belloche07,Bergin10}, and has been constrained by radiative transfer modeling of both single-dish \citep[e.g.][]{Hatchell03,Rolffs11apex} and interferometry observations \citep[e.g.][]{Osorio09}.

Especially interesting is the hottest molecular gas, which is closest to the heating sources. A good tracer is vibrationally excited HCN ($v_2$=1), whose rotationally excited levels lie $\sim 1000$ K higher in energy than the ground vibrational state, and which has transitions in the cm-wave regime, whose radiation is not obscured by dust. These so-called direct $\ell$-type  transitions are intrinsically weak, but have been detected in Galactic high-mass star-forming regions \citep{Thorwirth01} as well as toward the proto-planetary nebula CRL618 \citep{Thorwirth03} and the starburst galaxy Arp220 \citep{Salter08}.

Obviously, this hot gas is very compact, and high spatial resolution is needed to investigate it. In this paper, we present VLA observations (with among the highest resolution ever obtained for thermal molecular line emission) of three massive hot cores which have strong HCN direct $\ell$-type lines (as measured with the Effelsberg 100-m telescope, Sect.~\ref{sec:sd}). The sources are G10.47+0.03 \citep[e.g. ][]{Wyrowski99}, which lies at a distance of 10.6 kpc \citep{Pandian08} and has a luminosity of about $7\times 10^5$ L$_\odot$ \citep{Cesaroni10}, and SgrB2-N and -M near the Galactic center at 7.8 kpc distance \citep{Reid09}, which have total luminosities of about $8.4\times 10^5$ and  $6.3\times 10^6$ L$_\odot$, respectively \citep{Goldsmith92}. All sources are sites of very active high-mass star formation and contain several H{\sc ii} regions.

After describing observations and data reduction (Sect.~\ref{sec:obs}), we present the results (Sect.~\ref{sec:results}) and the radiative transfer modeling (Sect.~\ref{sec:modeling}), discuss the implications (Sect.~\ref{sec:discussion}) and draw conclusions (Sect.~\ref{sec:concl}).

\begin{table*}
\caption[]{Observational summary}
\label{tab:obs}
\begin{tabular}{l cc cc c c c }
\hline\hline
Source & \multicolumn{2}{c}{Pointing center} & Continuum beam\tablefootmark{a} & Line beam\tablefootmark{a} & rms 1~km~s$^{-1}$\tablefootmark{b} &  rms continuum\tablefootmark{b} &  rms integrated line\tablefootmark{b}\\
%Phase Calibrator &  Amplitude Calibrator & Velocity Coverage\\
~ & RA(J2000) & Dec(J2000) & ($''; \degr$) & ($''; \degr$)  & (mJy/Beam)& (K) & (K km~s$^{-1}$)\\
\hline
G10.47+0.03& 18 08 38.236 & -19 51 50.3 &  $0.12 \times 0.06; 82$ & $0.14 \times 0.1;  69$ & 2.3 & 45 & 900 \\
SgrB2-N  & 17 47 19.902  & -28 22 17.8 & $0.12 \times  0.08; 60$ & $0.15 \times 0.11; 51$ & 3.0 & 75 & 1300\\
SgrB2-M  & 17 47 20.202 & -28 23 05.3 & $0.12 \times  0.08; 63$ & $0.15 \times 0.11; 51$ & 3.3 & 95 & 1400\\
\hline
\end{tabular}
\tablefoot{
\tablefoottext{a}{Different weighting schemes were used for continuum and line images. Major and minor axis and position angle are given.}
\tablefoottext{b}{rms noise in 1~km~s$^{-1}$ channel, continuum and integrated line maps (Figs.~\ref{fig:g10_maps}, \ref{fig:b2n_maps}, and \ref{fig:b2m_maps}). The noise is evaluated over a box from $-2$ to $-1''$ in both RA and Dec relative to the phase center.}
}
\end{table*}

\section{Observations and data reduction}\label{sec:obs}

\subsection{VLA}

With the National Radio Astronomy Observatory's (NRAO's)\footnote{The National Radio Astronomy Observatory is a facility of the National Science Foundation operated under cooperative agreement by Associated Universities, Inc.} Very Large Array (VLA) in its BnA configuration we observed three massive star forming regions in the direct $\ell$-type line of vibrationally excited HCN at 40.7669 GHz (project AR687). This transition connects the two levels of the $v_2$=1, $J$=13 doublet, whose energies lie 1411 and 1413 K above the ground state, respectively. 
%located near Socorro, New Mexico. 

Each of the sources was observed during LST 15-21; G10.47+0.03 on 2009 January 30, SgrB2-M on January 31,  and SgrB2-N on February 1. Bandpass calibration was done on 3C279 and 3C454.3, and primary flux calibration on 3C286. The observations were performed in fast-switching mode with 150s on the source and 30s on a phase calibrator (1755-225 for G10.47+0.03 and SgrA* for SgrB2), yielding an on-source integration time of about 4.5 hours per source. For amplitude calibration, 1733-130 (G10.47+0.03) or SgrA* (SgrB2) was observed every 20 minutes, and reference pointing was done every hour on these sources.

For G10.47+0.03, we centered one IF with 6.25 MHz bandwidth at 40.762 GHz (shifted from the rest frequency due to the apparent source motion) and one with 25 MHz bandwidth at 40.78 GHz to increase the continuum sensitivity. For the SgrB2 sources, the line is broader, so we used two overlapping IFs with 6.25 MHz bandwidth centered at 40.761 GHz and 40.765 GHz. The channel width is 0.72 km s$^{-1}$. %This results in a usable bandwidth of about 40 km/s for G10.47+0.03 and 67 km/s for SgrB2-M and -N.

Calibration and editing was done in AIPS: Bad data points were flagged; phase, amplitude, and flux were calibrated using the continuum data of the calibrator sources, and bandpass calibration was applied. In MIRIAD\footnote{http://bima.astro.umd.edu/miriad} \citep{Sault95}, Doppler correction was done and the channels were averaged to a width of 1 km~s$^{-1}$, covering LSR velocities of 48--87 km~s$^{-1}$ in G10.47+0.03 and 27--93 km~s$^{-1}$ in SgrB2. The continuum was fitted using line-free channels (excluding 58--77, 44--80, and 42--82 km~s$^{-1}$ in G10.47+0.03, SgrB2-N, and SgrB2-M, respectively) and subtracted from the line data. Imaging was done with
almost uniform weighting for continuum maps (MIRIAD robust parameter -2, resulting in beams of $0.12'' \times 0.06''$ for G10.47 and $0.12'' \times 0.08''$ for SgrB2) and more natural weighting for line maps (robust 0.5, resulting in beams of $0.14'' \times 0.1''$ for G10.47 and $0.15'' \times 0.11 ''$ for SgrB2). For total fluxes and lower noise levels, the continuum was also imaged in natural weighting.

%different weigting schemes (resolutions between 0.06 and 0.17$''$)
A pixel size of 0.02$''$ and a cutoff for cleaning of twice the rms noise in the image was used. Table~\ref{tab:obs} summarizes beam sizes and noise levels.     %In the continuum images, it is 0.5, 1 and 1.3 mJy/Beam
%!  convolve nat.weighted images to .25" or .2"

 For G10.47+0.03, the continuum clean components were used to self-calibrate the phases and apply the solutions to the line data. The absorption lines in this source were almost invisible before self-calibration. The procedure did not improve the SgrB2 data, presumably due to the good calibration with SgrA* only 0.7$^\circ$ away. All figures shown in this paper were created with the GILDAS software\footnote{http://www.iram.fr/IRAMFR/GILDAS}.

\subsection{Effelsberg}

With the Effelsberg 100-m telescope of the Max-Planck-Institut f\"ur Radioastronomie, we observed several sources in HCN direct $\ell$-type lines in 2000 \citep{Thorwirth01}, 2002, and 2007. The lines were the $J$=9 transition at 20.18 GHz, $J$=10 at 24.66 GHz, $J$=11 at 29.58 GHz, and $J$=12 at 34.95 GHz. Beam sizes are 37, 31, 26, and 22$''$, respectively. Pointing was done on nearby quasars, and is accurate to about 10$''$. The different polarizations and scans were averaged, and a baseline was subtracted by fitting a polynomial of order 0 to 5 to line-free channels. The noise tube units measured by the telescope were converted to fluxes by comparison with a source of known flux, mostly NGC7027 \citep{Ott94}, which was observed once per day. To correct the deformation of the telescope at low elevations, the resulting flux was multiplied by an additional factor of $\left(1-0.01\times (30^\circ - elv)\right)^{-1}$ \citep{Gallimore01}, which is 1.25 at an elevation of $10^\circ$, typical for SgrB2.

The observations were complemented by data from the millimeter-wave line survey of SgrB2 with the IRAM 30-m telescope \citep{Belloche08}.

\section{Results}\label{sec:results}

\begin{figure}
  \centering
  \includegraphics[angle=0,width=0.5\textwidth]{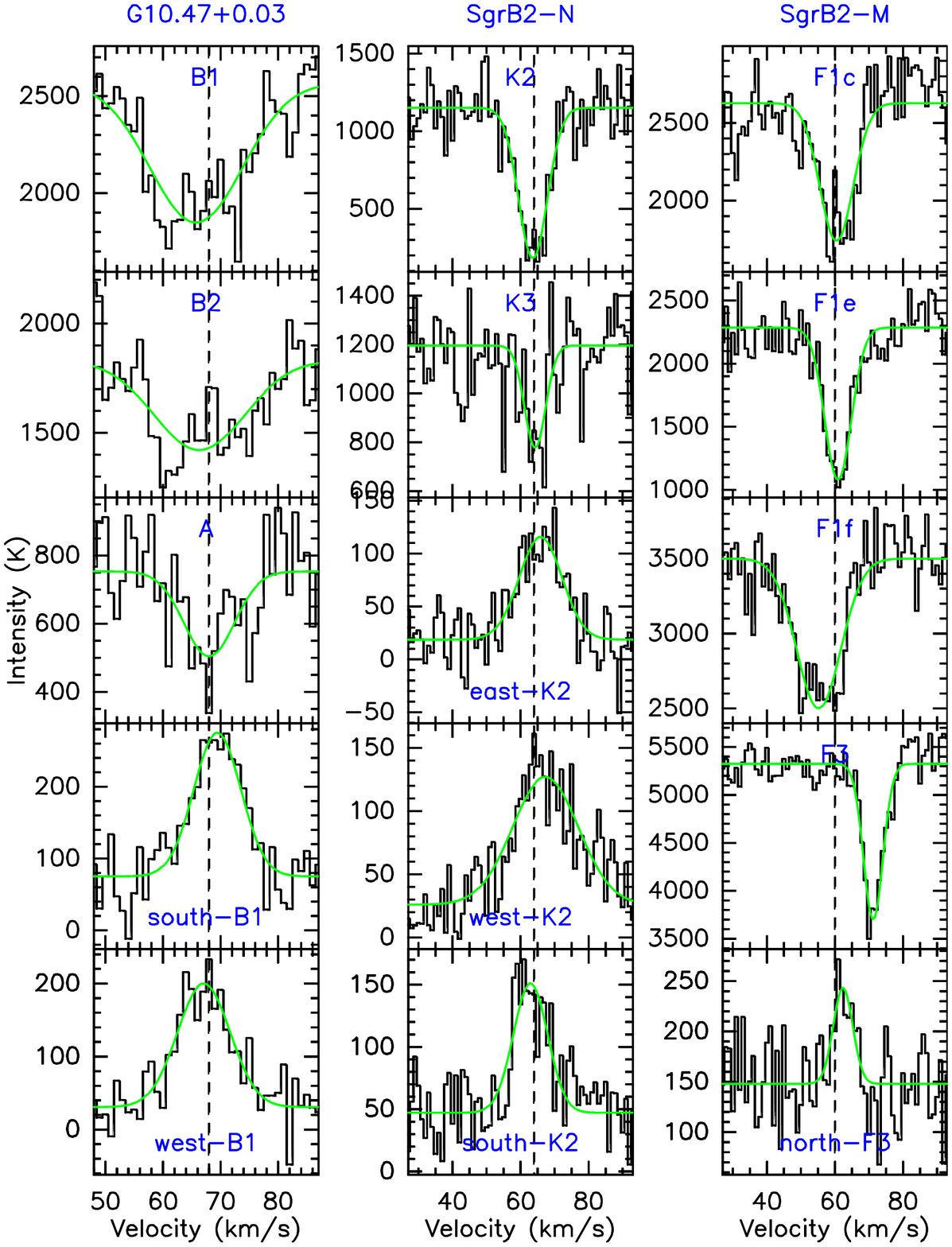} 
  \caption{The $v_2$=1, $J$=13 direct $\ell$-type line of HCN observed toward several positions in G10.47+0.03 (left column), SgrB2-N (central column), and SgrB2-M (right column). The emission lines are averaged over the areas shown in  Figs.~\ref{fig:g10_maps}, \ref{fig:b2n_maps},  and \ref{fig:b2m_maps}, while the absorption lines are observed with the beams given in Table~\ref{tab:obs}. Overlaid in green are Gaussian fits, whose parameters are given in Tables~\ref{tab:abslines} and \ref{tab:emlines}. The vertical dashed lines mark the assumed systemic velocities.}
  \label{fig:spectra}
\end{figure}

\begin{table*}
\caption[]{Parameters of absorption lines shown in Fig.~\ref{fig:spectra}}
\label{tab:abslines}
\begin{tabular}{l  cc   c  ccc cc }
\hline\hline
H{\sc ii} region & RA & Dec. & continuum\tablefootmark{a} &  line width\tablefootmark{b} & velocity & line/continuum & $\tau$ & HCN column density\tablefootmark{c} \\
 & (J2000) & (J2000) & (K) & (km~s$^{-1}$) & (km~s$^{-1}$) & & & (cm$^{-2}$) \\
\hline
G10.47+0.03 B1 & 18 08 38.243 & -19 51 50.22 & 2600 & 18.9 & 65.8 & 0.72  & 0.33 &  5.5(19)\\
G10.47+0.03 B2 & 18 08 38.22  & -19 51 50.62 & 1800 & 18.5 & 66.4 & 0.776 & 0.25 &  4.1(19) \\
G10.47+0.03 A  & 18 08 38.188 & -19 51 49.74 & 750  & 10.3 & 68   & 0.67  & 0.4   &  3.6(19) \\
\hline
SgrB2-N K2 & 17 47 19.882  & -28 22 18.45 & 1150 & 10.0 & 63.6 & 0.16  &  1.83  & 1.6(20) \\ %1.75 & 1.5(20)\\
SgrB2-N K3 & 17 47 19.902  & -28 22 17.18 & 1200 &  7.3 & 64.3 & 0.652 &  0.43 & 2.8(19) \\
\hline
SgrB2-M F1c & 17 47 20.126 & -28 23 03.9 & 2650 & 11.5 & 60.4 & 0.664 & 0.41  & 4.1(19) \\
SgrB2-M F1e & 17 47 20.131 & -28 23 04.02 & 2300 & 9.2 & 60.9 & 0.474 & 0.75 & 6.1(19) \\
SgrB2-M F1f & 17 47 20.147 & -28 23 03.88 & 3500 & 15.3 & 55.2 & 0.714 & 0.34 & 4.4(19) \\
%\multirow{3}{*}{SgrB2-M F1f} & \multirow{3}{*}{17 47 20.147} & \multirow{3}{*}{-28 23 03.88} & \multirow{3}{*}{3750} & 16.1 & 55.4 &  0.72 & 0.33 & 4.7(19)\\
% & & & & 11.3 & 51.6 & 0.763 & 0.27 & 2.7(19)\\
% & & & & 6.7 & 60.3 & 0.76 & 0.27 & 1.6(19)\\
SgrB2-M F3 & 17 47 20.172 & -28 23 04.58 & 5350 & 6.7 & 71.2 & 0.696 & 0.36 & 2.1(19) \\ %0.36 & 2(19)\\
\hline
\end{tabular}
\tablefoot{The beam size is $0.14'' \times 0.1''$ for G10.47+0.03 and $0.15'' \times 0.11''$ for SgrB2. The rms on 1 km~s$^{-1}$ channels is 120-150 K. Results of Gaussian fits and lower limits on the optical depth $\tau$ and on the column density of hot HCN are given. \tablefoottext{a}{The continuum flux density is the brightness temperature at the given locations in the line maps (which have a larger beam than the pure continuum maps). To convert from K to mJy/Beam, one has to divide by 53 for G10.47+0.03 and by 45 for SgrB2.} \tablefoottext{b}{Full Width at Half Maximum (FWHM) of the Gaussian} \tablefoottext{c}{The numbers in parentheses are powers of 10.}}
\end{table*}

\begin{table}
\caption[]{Results of Gaussian fits to the emission lines shown in Fig.~\ref{fig:spectra}}
\label{tab:emlines}
\begin{tabular}{l  ccc  }
\hline\hline
field &  line width\tablefootmark{a} & velocity & intensity\\
 &    (km~s$^{-1}$) & (km~s$^{-1}$) & (K)  \\
\hline
G10.47+0.03 south-B1  & 9.7  & 69.4 &  200 \\
G10.47+0.03 west-B1   & 10.8 & 67.1 &  170 \\
\hline
SgrB2-N east-K2   & 14.9 & 65.9 & 100 \\
SgrB2-N west-K2   & 22.9 & 67.3 & 100 \\
SgrB2-N south-K2  & 11.9 & 62.9 & 100  \\
\hline
SgrB2-M north-F3  & 6.8 & 62.4 & 100  \\
\hline
\end{tabular}
\tablefoot{
\tablefoottext{a}{Full Width at Half Maximum (FWHM) of the Gaussian}}
\end{table}

\begin{figure*}
  \centering
  \includegraphics[bb=40         28        534      640,angle=-90,width=0.49\textwidth]{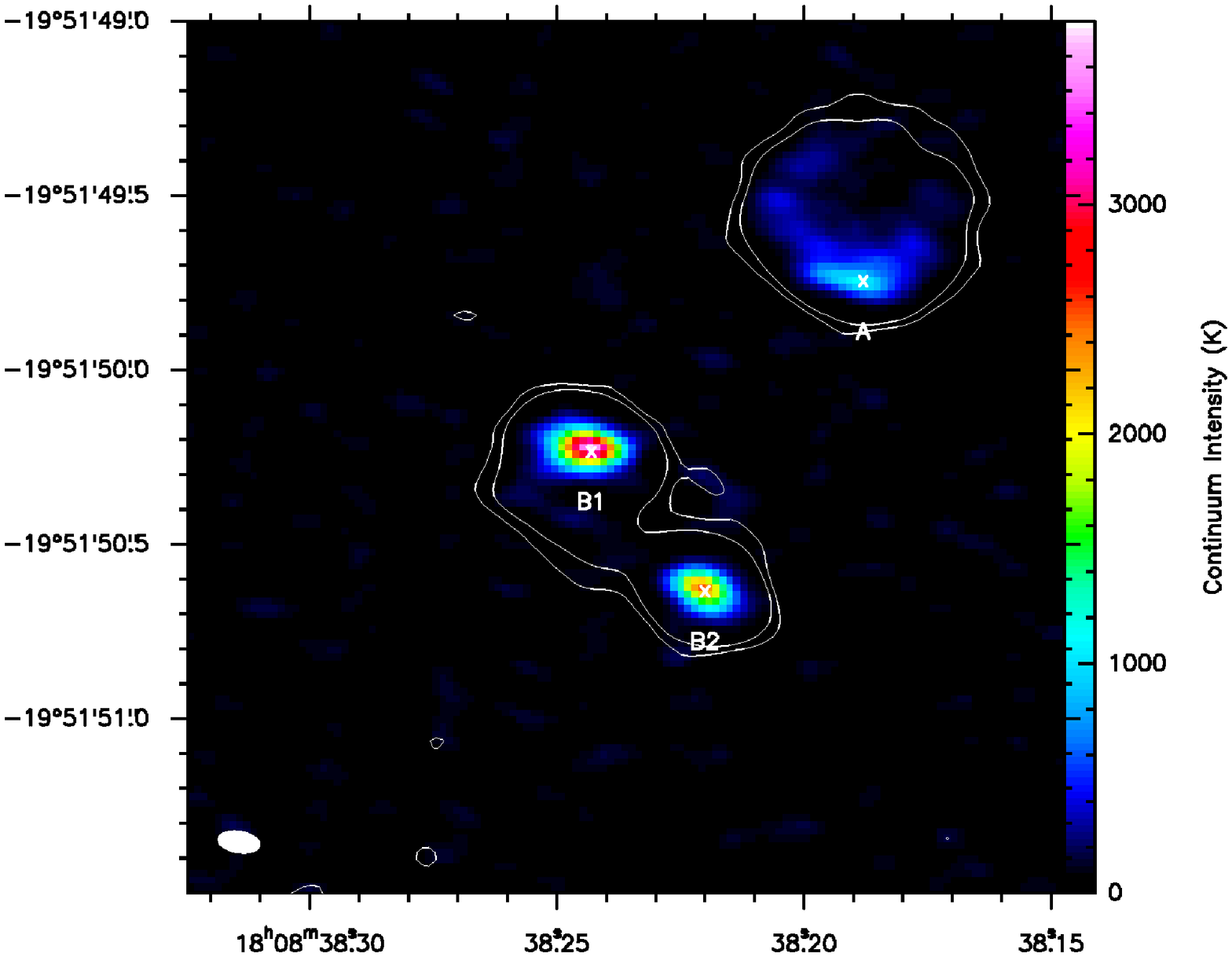} 
  \includegraphics[bb=40         28        534      640,angle=-90,width=0.49\textwidth]{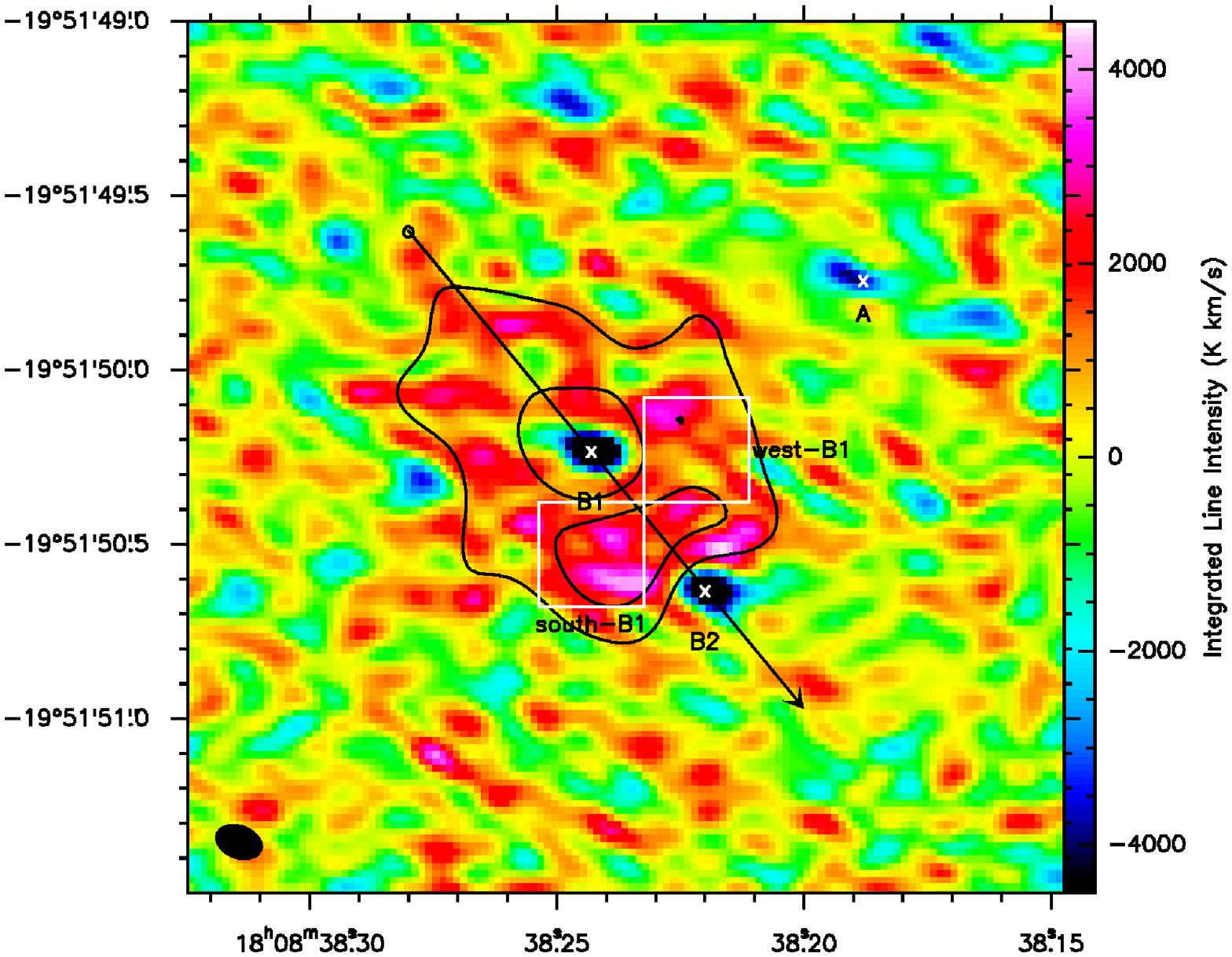} 
  \caption{VLA maps of G10.47+0.03. The 7mm continuum emission is shown on the left, the integrated line map on the right (beams are depicted in the lower left). Contours denote 3 and 6 $\sigma$ levels of the naturally weighted continuum (white, left) and of the integrated line emission convolved to 0.25$''$ resolution (black, right). The white crosses give the positions toward which the absorption spectra were obtained and the white boxes the area over which the emission spectra are integrated, shown in Fig.~\ref{fig:spectra}. The arrow denotes the direction along which the position-velocity plot of Fig.~\ref{fig:g10_pv} was obtained. The map size  is 2.5$''$, about 0.13 pc.}
  \label{fig:g10_maps}
\end{figure*}

%In this section we present the results of our observations. 
For each source we show spectra (Fig.~\ref{fig:spectra}), continuum and integrated line maps, as well as position-velocity diagrams. Absorption lines are summarized in Table~\ref{tab:abslines} and emission lines in Table~\ref{tab:emlines}. 

From Gaussian fits to the absorption lines, lower limits on the optical depth $\tau$ and on the column density of hot HCN can be derived: The line-to-continuum ratio is $e^{-\tau}$ if line emission can be neglected, i.e. if the beam-averaged (excitation) temperature of the absorbing gas is much lower than the beam-averaged background brightness temperature - else the line-to-continuum ratio is larger than $e^{-\tau}$. The background radiation is provided by free-free emission from ionized gas of around 10$^4$ K, which is at least an order of magnitude larger than the temperature of the absorbing gas. However, the observed continuum is lower than 10$^4$ K due to optically thin free-free radiation in parts of the beam (see e.g. \citet{Cesaroni10} for spectral energy distributions of H{\sc ii} regions with density gradient). Additionally, the hot molecular gas could emit over the whole beam. Therefore, the real optical depths can be much higher than derived from the line-to-continuum ratio, and the values given in Table~\ref{tab:abslines} are only lower limits. 

From $\tau$ and the observed line width $\Delta v$ (FWHM), the HCN column density can be derived, assuming Local Thermodynamic Equilibrium (LTE, see also Sect.~\ref{sec:assumptions}) at a temperature $T$:  Using 
\begin{equation}\label{eq:tau}
\tau = \frac{c^2}{8\pi\nu^2} N_{\rm u} A_{\rm ul} \Phi \left( e^{\frac{h\nu}{kT}}-1\right)
\end{equation}
 with frequency $\nu=40.7669$ GHz, Einstein A coefficient $A_{\rm ul}=3.75\times 10^{-8}$ s$^{-1}$ and profile function at Gaussian line center $\Phi= \sqrt{\frac{ln(2)}{\pi}} \times \frac{2}{\Delta v} \approx \frac{0.94}{\Delta v}$, the total column density $N$ can be obtained from the column density in the upper state 
\begin{equation}\label{eq:N}
N_{\rm u}=N \frac{g_{\rm u}}{Q(T)}e^{-\frac{E_u}{kT}}
\end{equation}
inserting the statistical weight $g_{\rm u}=2J+1=27$, the energy of the upper level $E_{\rm u} = 1413$ K, and the partition function $Q(T)$ \citep{HCN_Q}. The temperature has two competing effects on these equations: While a higher temperature raises $N_{\rm u}$, it lowers the population difference between upper and lower level (last factor of Equation~\ref{eq:tau}). For given $\tau$ and $\Delta v$, a minimum of $N$ is reached around 500 K, while twice as many molecules are required at around 300 and 1000 K, and 10 times as many at 200 and 1800 K (100 times at 140 K). In Table~\ref{tab:abslines}, we give this minimum, which can be regarded as a lower limit on the column density of hot HCN.  Only if the level population difference is higher than expected from LTE, the true HCN column density would be lower.

%Assuming an optical depth of 0.5 for the emission lines, the temperature is around K
%The integrated line emission

\subsection{G10.47+0.03}

\begin{figure}
  \centering
  \includegraphics[bb=98        243        540        733,angle=-90,width=0.49\textwidth]{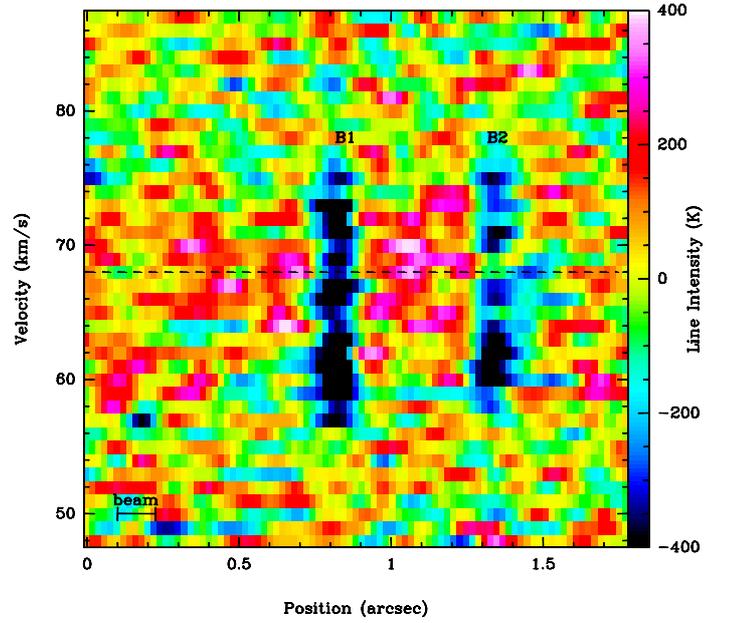} 
  \caption{Position-velocity diagram of G10.47+0.03, along the arrow shown in Fig.~\ref{fig:g10_maps}. The velocity of 68 km~s$^{-1}$ is marked as a dashed line, and the color scale is centered at 0 K. }
  \label{fig:g10_pv}
\end{figure}

The 7mm continuum (Fig.~\ref{fig:g10_maps} left) shows the two hypercompact H{\sc ii} regions B1 and B2 as well as the cometary-shaped ultracompact H{\sc ii} region A \citep[see also ][]{Cesaroni10}. The total fluxes are 80 mJy for B1, 48 mJy for B2, and 79 mJy for A.

The integrated line map (Fig.~\ref{fig:g10_maps} right) shows the absorption and the distribution of line emission, which is concentrated around B1 and especially strong between B1 and B2. The absorption lines toward B1 and B2 have a component at lower velocities than the systemic velocity of around 68 km~s$^{-1}$ (derived from emission lines), indicating expansion motions (Figs.~\ref{fig:spectra} and \ref{fig:g10_pv}).

\subsection{SgrB2-N}

\begin{figure*}
  \centering
  \includegraphics[bb=40         28        534      640,angle=-90,width=0.49\textwidth]{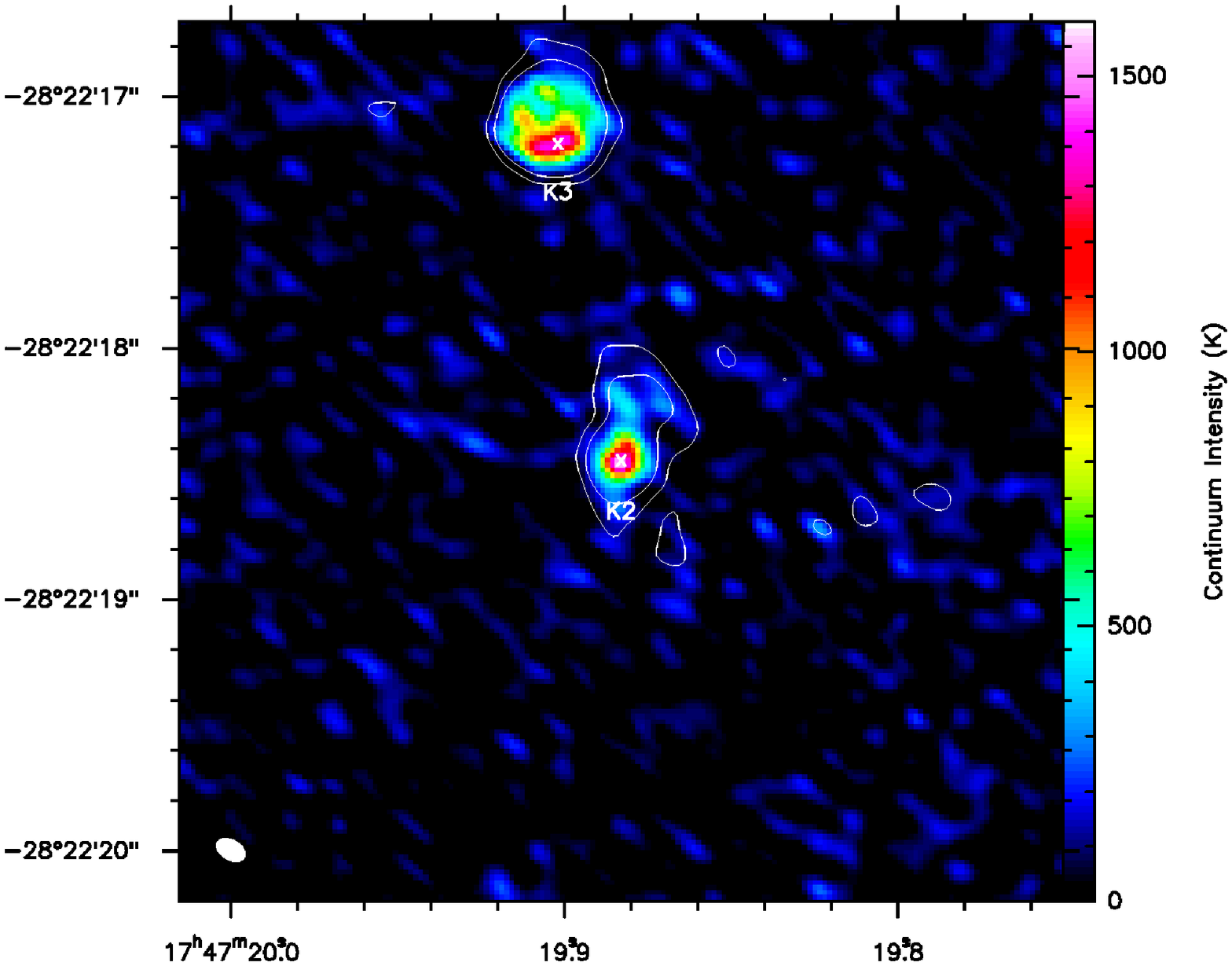} 
  \includegraphics[bb=40         28        534      640,angle=-90,width=0.49\textwidth]{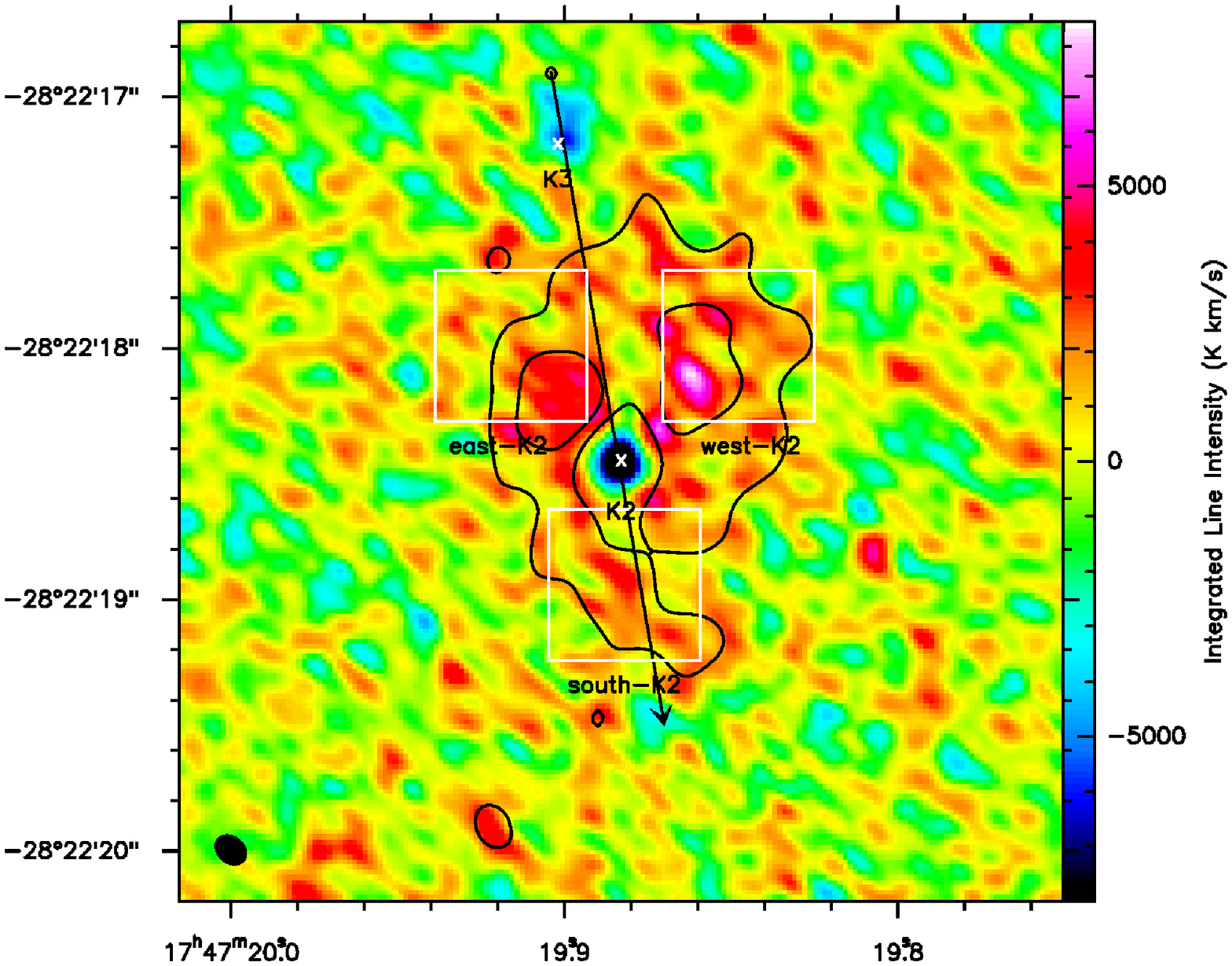} 
  \caption{VLA maps of SgrB2-N.  The 7mm continuum emission is shown on the left, the integrated line map on the right (beams are depicted in the lower left). Contours denote 3 and 6 $\sigma$ levels of the naturally weighted continuum (white, left) and of the integrated line emission convolved to 0.25$''$ resolution (black, right). The white crosses give  the positions toward which the absorption spectra were obtained  and the white boxes the area over which the emission spectra are integrated, shown in Fig.~\ref{fig:spectra}. The arrow denotes the direction along which the position-velocity plot of Fig.~\ref{fig:b2n_pv} was obtained. The map size  is 3.5$''$, about 0.13 pc. }
  \label{fig:b2n_maps}
\end{figure*}

\begin{figure}
  \centering
  \includegraphics[bb=98        243        540        733,angle=-90,width=0.49\textwidth]{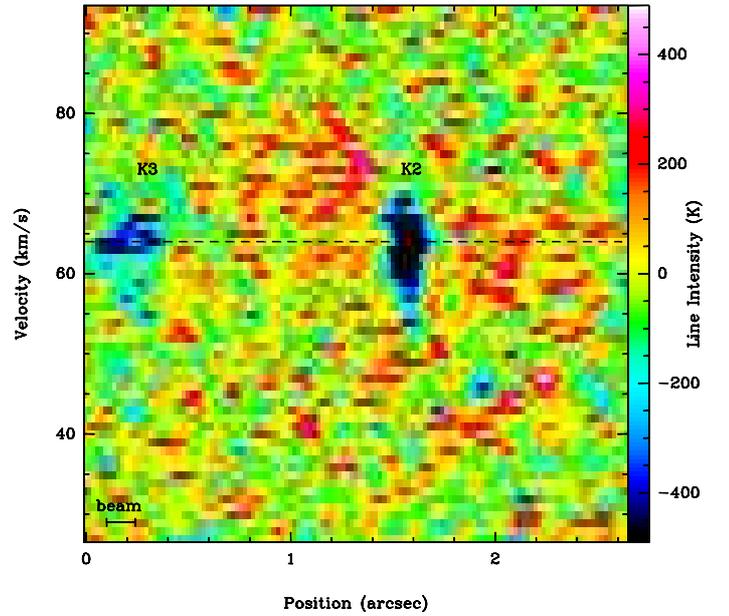} 
  \caption{Position-velocity diagram of SgrB2-N, along the arrow shown in Fig.~\ref{fig:b2n_maps}. The velocity of 64 km~s$^{-1}$ is marked as a dashed line, and the color scale is centered at 0 K. }
  \label{fig:b2n_pv}
\end{figure}

The continuum map (Fig.~\ref{fig:b2n_maps} left) shows K2 and K3 \citep{Gaume95} with fluxes of 70 and 140 mJy, respectively. Outside of the range shown in Fig.~\ref{fig:b2n_maps}, we also detect K4, K1 (as a north-south elongated, 0.5$''$ long H{\sc ii} region of around 100 mJy), %at R.A. 17:47:19.784, Dec. -28:22:20.5 with a total flux of 80 mJy) 
and a 10 mJy point source at R.A. 17:47:19.905, Dec. -28:22:13.46 (4$''$ north of K3). This latter source is associated with C$_2$H$_5$CN emission at 75 km~s$^{-1}$ and 3mm continuum detected by \citet{LiuSnyder99}. We note that emission from complex molecules \citep[such as amino acetonitrile, ][]{Belloche08}  peaks at K2.

The integrated line map (Fig.~\ref{fig:b2n_maps} right) shows wide-spread line emission around K2. Toward K2 itself, the line absorbs over 80\% of the continuum, implying an enormous optical depth (Fig.~\ref{fig:spectra} and Table~\ref{tab:abslines}). The southern part seems to have a bit lower velocities than the northern part (Figs.~\ref{fig:spectra} and ~\ref{fig:b2n_pv}).

\subsection{SgrB2-M}

\begin{figure*}
  \centering
  \includegraphics[bb=40         28        534      640,angle=-90,width=0.49\textwidth]{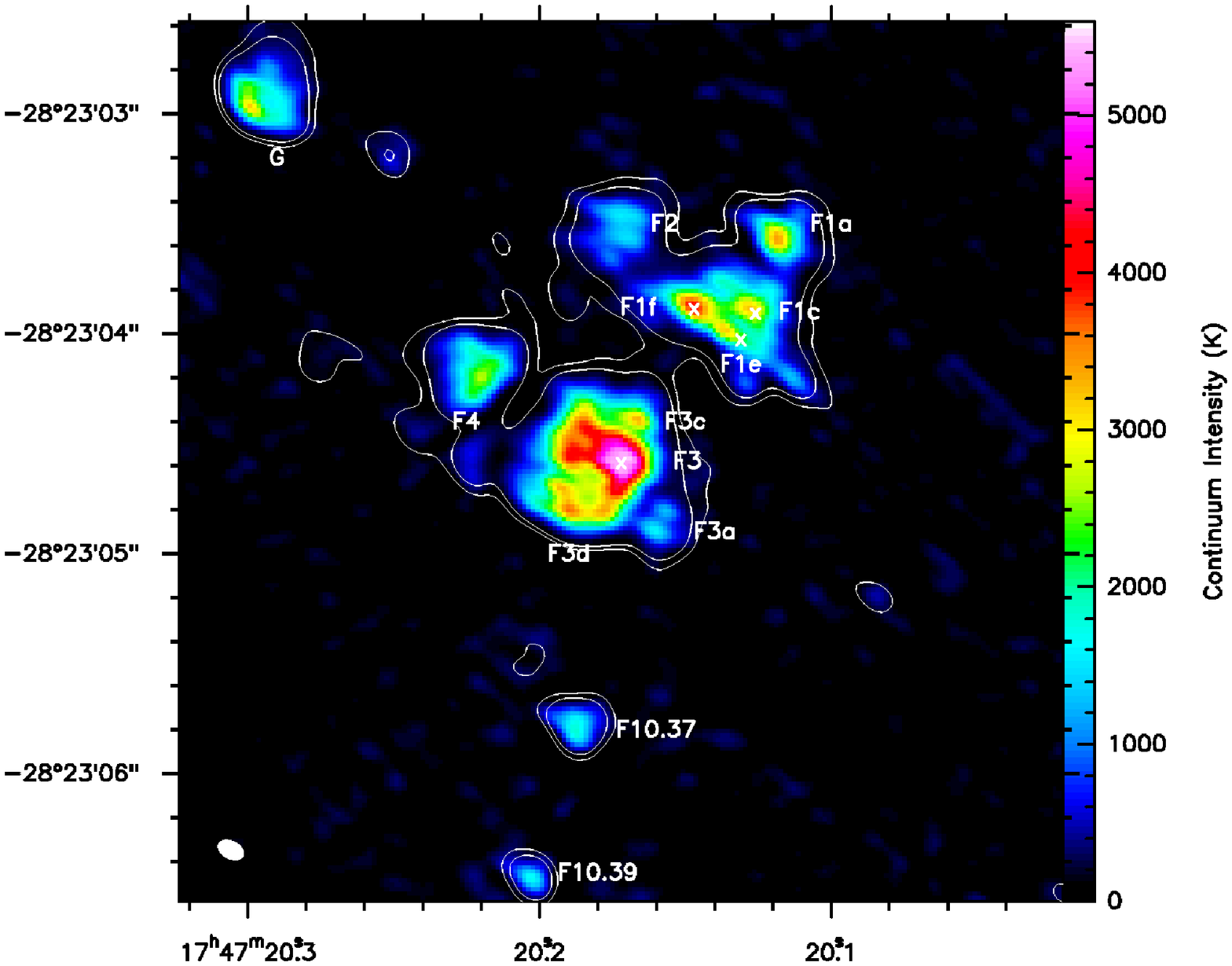} 
  \includegraphics[bb=40         28        534      640,angle=-90,width=0.49\textwidth]{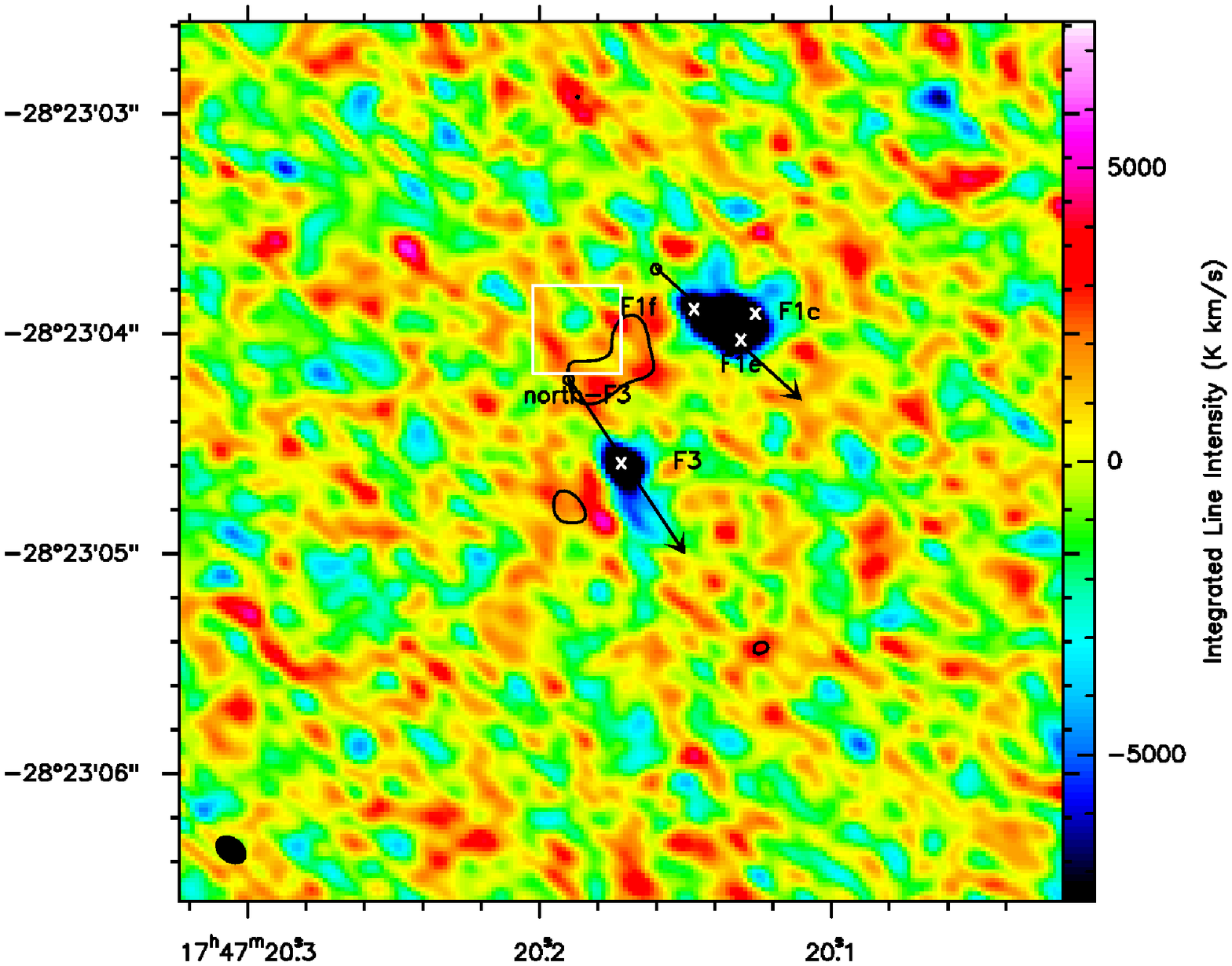} 
  \caption{VLA maps of SgrB2-M. The 7mm continuum emission is shown on the left, the integrated line map on the right (beams are depicted in the lower left). Contours denote 3 and 6 $\sigma$ levels of the naturally weighted continuum (white, left) and of the integrated line emission convolved to 0.25$''$ resolution (black, right). The white crosses give the positions toward which the absorption spectra were obtained and the white boxes the area over which the emission spectra are integrated, shown in Fig.~\ref{fig:spectra}. The arrow denotes the direction along which the position-velocity plot of Fig.~\ref{fig:b2m_pv} was obtained. The map size is 4$''$, about 0.15 pc. }
  \label{fig:b2m_maps}
\end{figure*}

\begin{figure*}
  \centering
  \includegraphics[bb=98        243        540        733,angle=-90,width=0.49\textwidth]{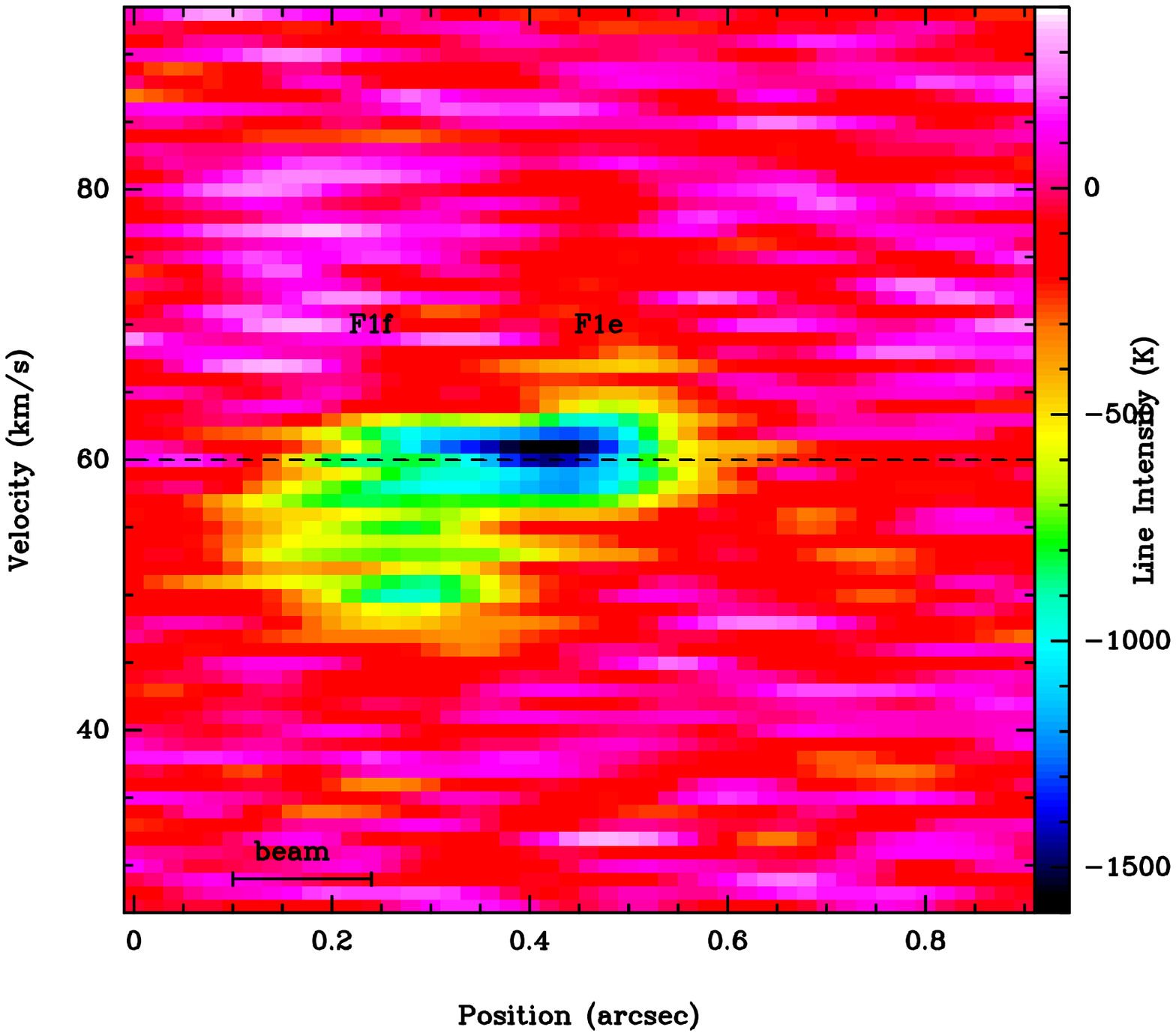} 
  \includegraphics[bb=98        243        540        733,angle=-90,width=0.49\textwidth]{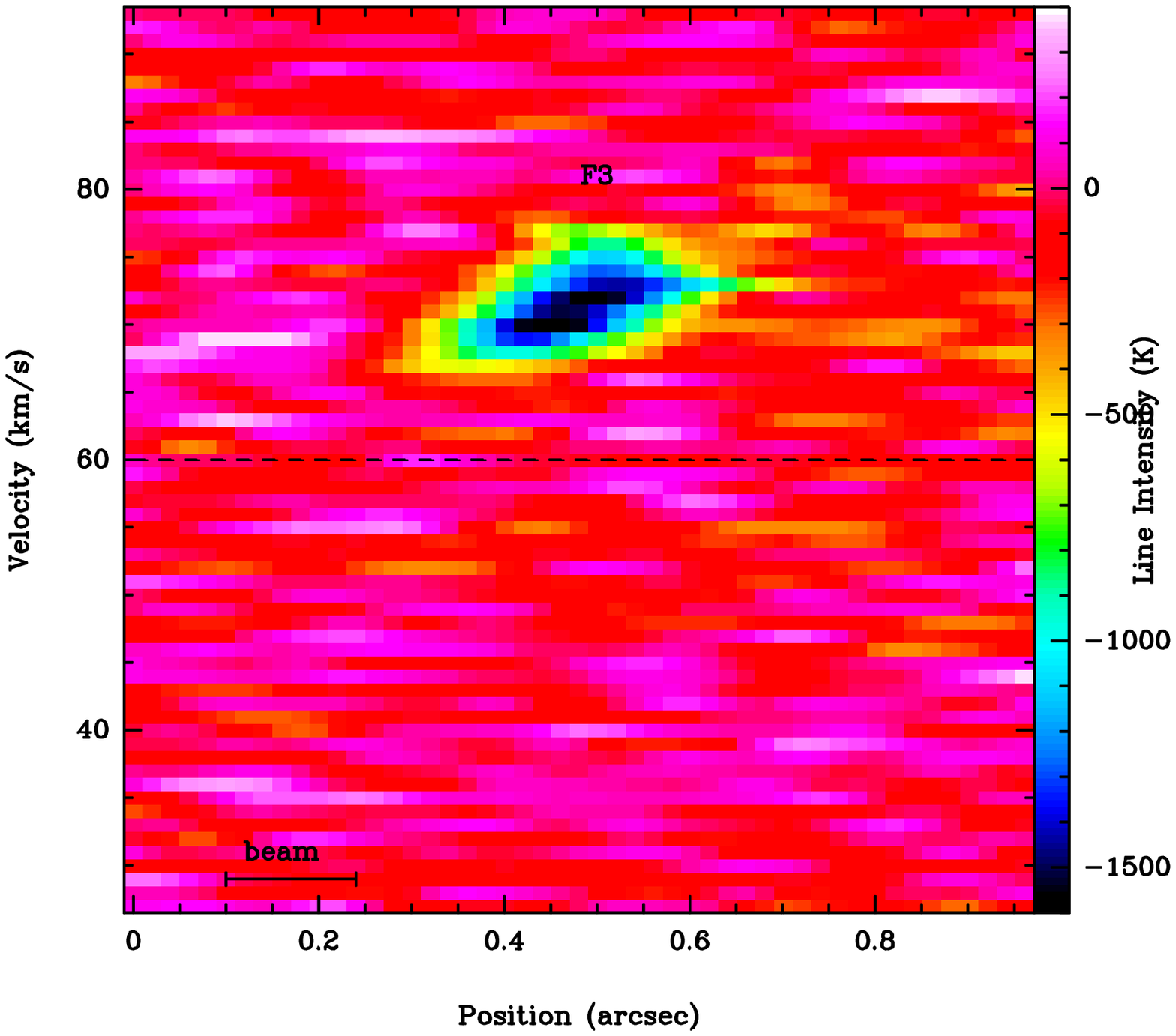} 
  \caption{Position-velocity diagrams of SgrB2-M, along the arrows shown in Fig.~\ref{fig:b2m_maps}. The cut through F1 is shown on the left, F3 on the right. The velocity of 60 km~s$^{-1}$ is marked as a dashed line, and the color scale is dominated by absorption. }
  \label{fig:b2m_pv}
\end{figure*}

The continuum map (Fig.~\ref{fig:b2m_maps} left) resembles the one obtained by \citet{dePree98}. This source consists of a whole cluster of hypercompact H{\sc ii} regions, including F3 with 1.3 Jy and F1c/e/f with about 0.5 Jy.

Line absorption is concentrated toward the strongest sources F1 and F3 (Fig.~\ref{fig:b2m_maps} right). Toward F1, the line is at 60 km~s$^{-1}$ with a component at 50 km~s$^{-1}$ in F1f (Fig.~\ref{fig:b2m_pv} left). A northeast-southwest velocity gradient from 69 to 73 km~s$^{-1}$ over 0.3$''$ (2300 AU) is detected in F3 (Fig.~\ref{fig:b2m_pv} right). In addition to the lines from Fig.~\ref{fig:spectra}, there is a possible absorption line toward F2 at 57 km~s$^{-1}$ with a width of 7 km~s$^{-1}$ and a line-to-continuum ratio of 0.8. Although not visible in the integrated line map, there is also line emission present, as seen by averaging over a field outside the H{\sc ii} regions.

\subsection{Single-dish data}\label{sec:sd}

\begin{figure}
  \centering
  \includegraphics[bb=58         34        453        744,angle=0,width=0.4\textwidth]{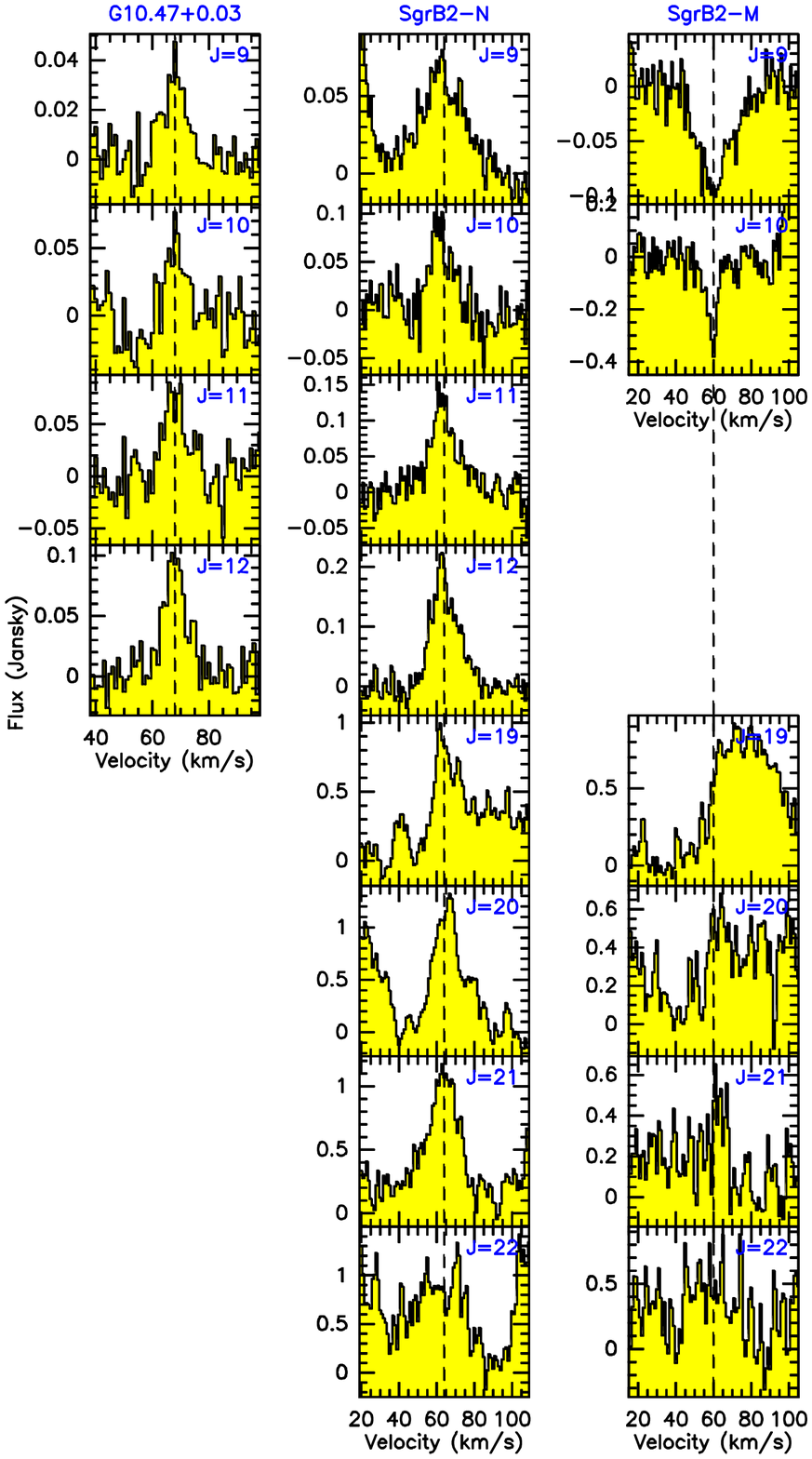} 
  \caption{Single-dish spectra of G10.47+0.03 (left), SgrB2-N (central), and SgrB2-M (right). $J$=9 (20 GHz) to 12 (35 GHz) were observed with the Effelsberg 100-m telescope, $J$=19 (85 GHz) to 22 (113 GHz) with the IRAM 30-m telescope. Beam sizes vary between 20 and 37$''$,  so the whole flux is received by the telescope.}
  \label{fig:sd}
\end{figure}

\begin{figure}
  \centering
  \includegraphics[bb= 58        535        453        660,angle=0,width=0.4\textwidth]{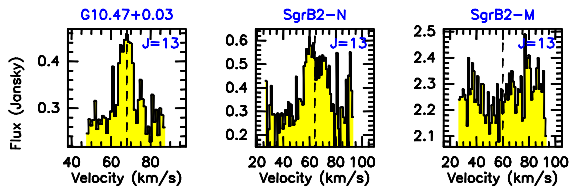} 
  \caption{VLA line flux of G10.47+0.03 (left), SgrB2-N (central), and SgrB2-M (right), integrated over the whole maps shown in Figs.~\ref{fig:g10_maps}, \ref{fig:b2n_maps}, and \ref{fig:b2m_maps}. }
  \label{fig:lineflux}
\end{figure}

\begin{figure}
  \centering
  \includegraphics[angle=-90,width=0.48\textwidth]{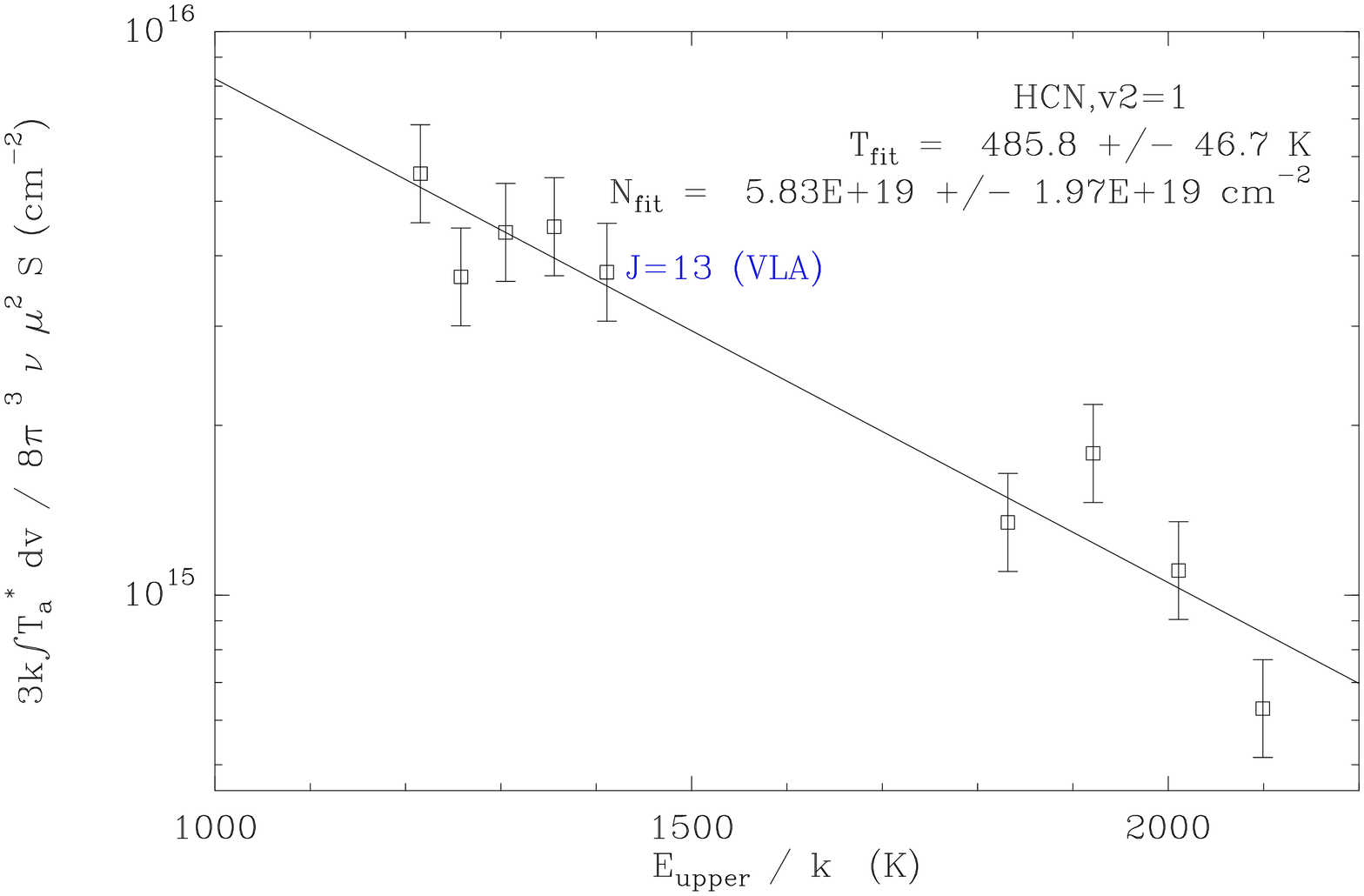} 
  \caption{Rotation diagram for SgrB2-N, based on the single-dish fluxes. The integrated line flux is source-averaged for a source size of 1$''$. The first four data points are from Effelsberg, the last four from IRAM 30-m; the VLA flux fits very well to the expectation.}
  \label{fig:rd}
\end{figure}

We investigated several sources with the Effelsberg 100-m telescope for HCN direct $\ell$-type transitions, with no detections in G31.41+0.31, G34.26+0.15, W3(H$_2$O), and W51d. Toward W51e, the $J$=12 line was detected (with half the intensity of G10.47+0.03), but the lower-$J$ transitions are questionable ($J$=9 at least 10 times lower than in G10.47+0.03). Clear detections in Galactic star-forming regions were only made in Orion-KL, G10.47+0.03, SgrB2-N, and -M. These latter three sources are shown in Fig.~\ref{fig:sd}, with additional IRAM 30-m data for SgrB2. $J$=19 is contaminated by an H$\gamma$ recombination line whose frequency corresponds to a $\sim$20 km~s$^{-1}$ higher velocity. $J$=20 is blended with a transition of ethanol in SgrB2-N and possibly with He$\gamma$ in -M.

To compare our VLA data to the single-dish data, Fig.~\ref{fig:lineflux} shows the line integrated over the whole map of Figs.~\ref{fig:g10_maps}, \ref{fig:b2n_maps}, and \ref{fig:b2m_maps}. The VLA flux in G10.47+0.03 and SgrB2-N agrees well with the expectations from the Effelsberg data. In SgrB2-M, absorption dominates at low frequencies, but in the IRAM 30-m data weak emission can be seen. Emission also contributes to the VLA flux.  

A bit puzzling is the different widths of $J$=9 and 10 in the Effelsberg data of SgrB2-M: It is 20 km~s$^{-1}$ in $J$=9 and only 6.6 km~s$^{-1}$ in $J$=10 (Fig.~\ref{fig:sd}). As we found no blending at the $J$=9 frequency, this is probably due to different ratios of the 50, 60, and 70 km~s$^{-1}$ components:  For $J$=10, the 60 km~s$^{-1}$ component dominates.

%Results of Gaussian fits are given in Table~\ref{tab:int}. In G10.47+0.03, the VLA flux agrees well with the expectations from the Effelsberg data, while the line flux of SgrB2-N is 20\% lower than predicted by the rotation diagram.  

In SgrB2-N, also the higher-$J$ transitions with level energies up to 2100 K were clearly detected. This allows to construct a rotation diagram (Fig.~\ref{fig:rd}), yielding a temperature of $485\pm 50$ K and an HCN column density in a 1$''$ source of $(5.8\pm 2) \times 10^{19}$ cm$^{-2}$. Note that taking absorption and optical depth into account would lead to a higher column density and a lower temperature, since both effects weaken especially the lower-$J$ lines.

%Using the models described in Section~\ref{sec:modeling}, we computed the $J$=12 line, yielding 0.15 Jy for G10.47+0.03 and  0.22 Jy for SgrB2-N, as compared to the measured values of 0.1 Jy for G10.47+0.03  and 0.2 Jy for SgrB2-N with the Effelsberg telescope.

% 
% \begin{table}
% \caption[]{Gaussian fits to the line integrated over the whole map shown in Figs.~\ref{fig:g10_maps}, \ref{fig:b2n_maps} and \ref{fig:b2m_maps}}
% \label{tab:int}
% \begin{tabular}{l c c ccc  }
% \hline\hline
% source  &  rms 1~km/s & continuum & line width & velocity & line intensity\\
%  &         (Jy)    & (Jy) & (km/s) & (km/s) & (Jy)  \\
% \hline
% G10.47+0.03 &  &0.21 & 7.3 & 67.6  & 0.15 \\
% 
% SgrB2-N &    &0.21 &  14.6 & 63.1   & 0.18 \\
% 
% SgrB2-M &    &1.9 &  14 & 58.5   & -0.13 \\
% 
% \hline
% \end{tabular}
% \end{table}
% 

%G10.47 $2 \times 2''$ (centered B1): continuum 215 mJy, line 140 mJy, width 7.1 km/s, vLSR 67.8 km/s
%B2N  $2 \times 3''$ (north-south elongated, centered K2): continuum 210 mJy, line 200 mJy, width 14.7 km/s, vLSR 63.5 km/s

\section{Modeling}\label{sec:modeling}

In an attempt to constrain the spatial structure of the hot molecular gas, we constructed radiative-transfer models that reproduce the observations. We employed a trial-and-error technique and compared the model to the data by eye. This method yields (at best) a model that is consistent with the data, but it cannot provide errors of the parameters, and different models could fit as well. The three-dimensional radiative-transfer code RADMC-3D\footnote{http://www.ita.uni-heidelberg.de/\textasciitilde dullemond/software/radmc-3d} was used to compute the radiation that a model emits. It allows adaptive mesh refinement, following user-defined criteria. Dust properties and the dust density distribution as well as  location, surface temperature and radius of stars are given as input. The program then computes the dust temperature by tracing photon packets which are randomly emitted by the stars. Additional inputs are density and temperature of ionized gas, the velocity field, fundamental molecular data, and the molecular abundance.

\subsection{Assumptions}\label{sec:assumptions}

We use the dust opacity from \citet{Ossenkopf94} without grain mantles or coagulation. This is reasonable assuming that ice mantles around dust grains have evaporated completely and recently, so that the grains have had no time to coagulate again. Since the VLA continuum is dominated by free-free emission, the data are not sensitive to the dust emission and so we cannot constrain the dust optical depth, which is a combination of dust opacity and density, through the dust emission. However, the heating is strongly affected by the dust optical depth.

The temperature of the ionized gas is assumed to be $10^4$ K, which is the order of magnitude expected from cooling by trace species. A different temperature would require a different electron density to account for the observed fluxes. 

The HCN molecular data are from \citet{HCN_rot_2003} and taken from the Cologne Database for Molecular Spectroscopy \citep{Mueller01,CDMS2_2005}. We use an HCN abundance of $10^{-5}$ relative to H$_2$ for our models. Since we mainly constrain the density of HCN, a different abundance would require different dust densities (with the gas/dust mass ratio of 100). Dust continuum data (from the SMA) indicate that the resulting dust densities are on the right order of magnitude, which justifies this abundance assumption.

The line radiative transfer assumes LTE (full non-LTE radiative transfer is also planned for RADMC-3D). LTE is a reasonable assumption for the observed line as HCN thermalizes to the ambient dust temperature: It is vibrationally excited by 14$\mu$m radiation emitted by warm dust, which is very optically thick at this wavelength due to the high densities (an optical depth of 1 is reached after about 25 AU in a density of $10^8$ cm$^{-3}$). In this warm and dense environment, also the levels in the ground vibrational state are thermalized by infrared pumping as well as by collisions and radiative excitation. Deviations from LTE occur at low temperatures, where the vibrational levels are not populated anyway.

The models have no macroscopic velocity field. We assume a constant intrinsic line width (microturbulence) throughout the source, which reflects internal motions. The resulting spectra are shifted by the source velocity (68 km~s$^{-1}$ for G10.47+0.03, 64 km~s$^{-1}$ for SgrB2-N, 61 km~s$^{-1}$ for SgrB2-M F1e, and 71 km~s$^{-1}$ for SgrB2-M F3).

\subsection{Modeling results}

To compare the model to the observational data, the synthetic map produced by RADMC-3D, supplemented by distance and coordinate information, is Fourier-transformed, folded with the uv coverage of the observations, and imaged again. The first step of our fitting procedure is to reproduce the observed continuum by creating  H{\sc ii} regions whose sizes and densities are based on \citet{Cesaroni10} for G10.47+0.03 and \citet{dePree98} for SgrB2-M. For SgrB2-N, we use similar models as for G10.47+0.03 since the 1.3cm data of \citet{Gaume95} are not sufficient in terms of angular resolution and wavelength. Sizes and densities of the H{\sc ii} regions are slightly modified to fit the observed continuum. The hypercompact H{\sc ii} regions B1 and B2 in G10.47+0.03 and K2 in SgrB2-N have thus density gradients (with $r^{-1.5}$ and outer radii of 1590 AU in B1, 950 AU in B2 and 1000 AU in K2), while all other H{\sc ii} regions have constant densities.  

\begin{figure*}
  \centering
  \includegraphics[bb=178         29        561        450,angle=-90,width=0.8\textwidth]{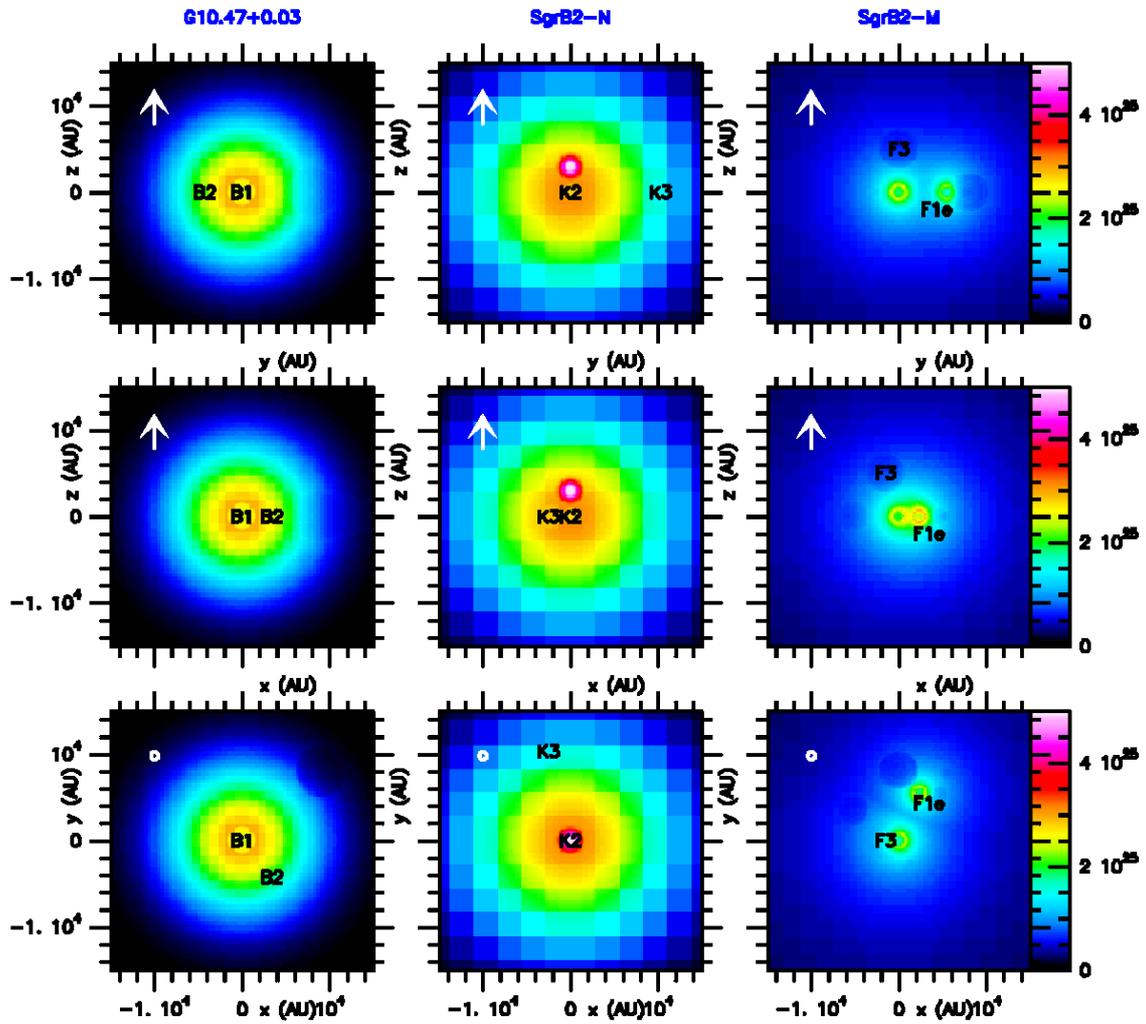}  
  \caption{Column density maps of the models for the three sources. The lower panels show the column density (in H$_2$ cm$^{-2}$) in z-direction, the central panels in y-direction, and the upper panels in x-direction (the arrow points to the observer, who is at positive z). Two H{\sc ii} regions are marked for each source. The model radiation is compared to the observations in Figs.~\ref{fig:g10_mod} to \ref{fig:b2m_mod}.}
  \label{fig:coldens}
\end{figure*}

\citet{Cesaroni10} and \citet{dePree98} also give spectral types of the exciting stars, which are converted to luminosity following \citet{Panagia73}. These stars heat the dust and are, for simplicity, placed in the plane of the sky, except for two H{\sc ii} regions in SgrB2-M. The dust density distribution is adapted to fit the line observations. Figure~\ref{fig:coldens} shows the column densities in x, y, and z directions of the three models that are described in the following.

\begin{figure*}
  \centering
  \includegraphics[angle=-90,width=0.8\textwidth]{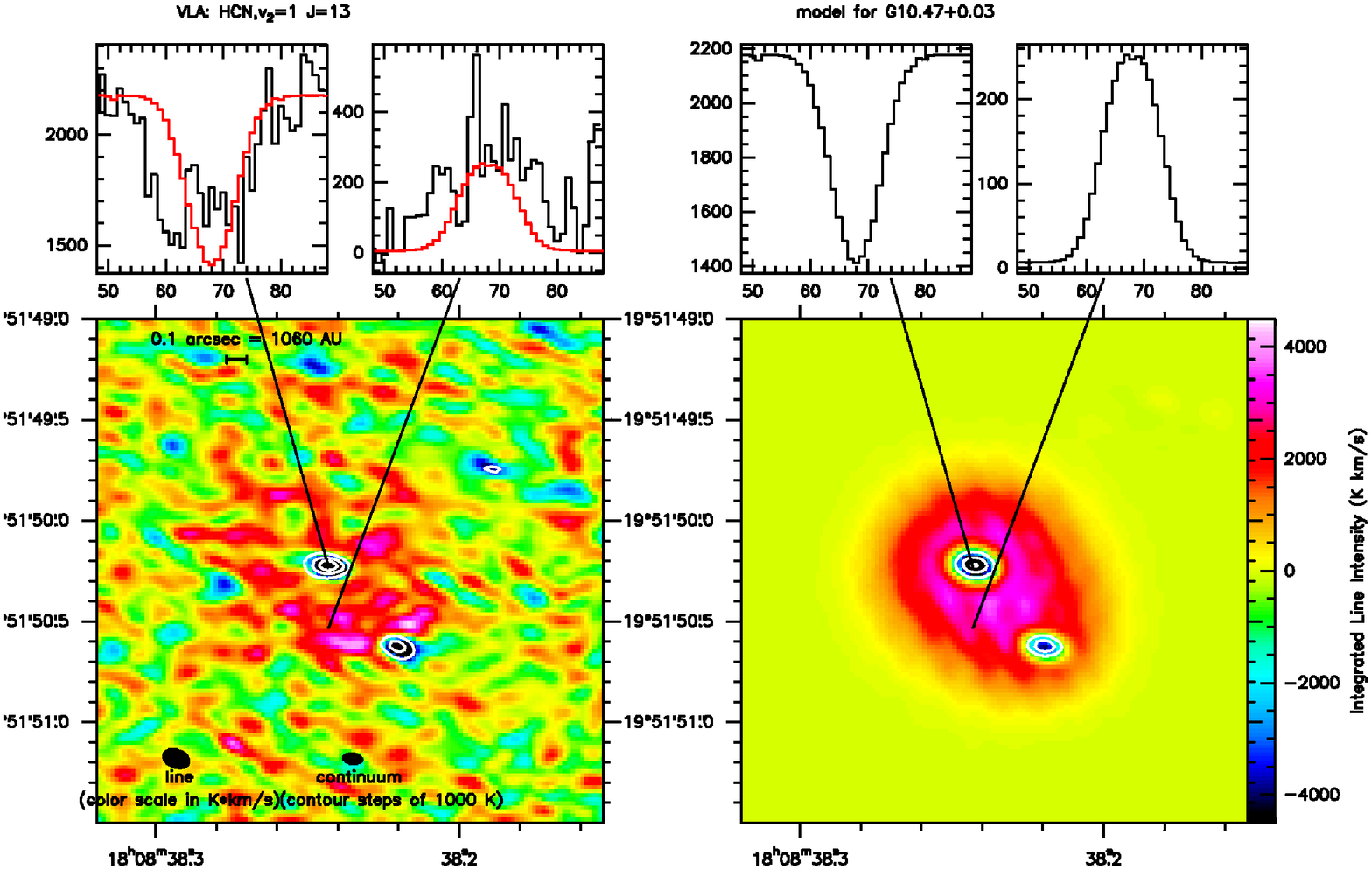}  
  \caption{Model for G10.47+0.03 (right) compared to the data (left). The model spectra are also overlaid in red. The white contours denote the continuum in steps of 1000 K.}
  \label{fig:g10_mod}
\end{figure*}

\paragraph{G10.47+0.03} The model consists of a clump centered at B1 whose density follows a Gaussian with $7\times 10^7$ H$_2$ cm$^{-3}$ at the half-power radius of 7000 AU. This dust is heated by B1 with $10^5$ L$_\odot$, B2 with $8.3\times 10^4$ L$_\odot$, and A with $4.6\times 10^4$ L$_\odot$. The intrinsic line width (FWHM) is 8.3 km~s$^{-1}$. The model output is compared to the observations in Fig.~\ref{fig:g10_mod}.

\begin{figure*}
  \centering
  \includegraphics[angle=-90,width=0.8\textwidth]{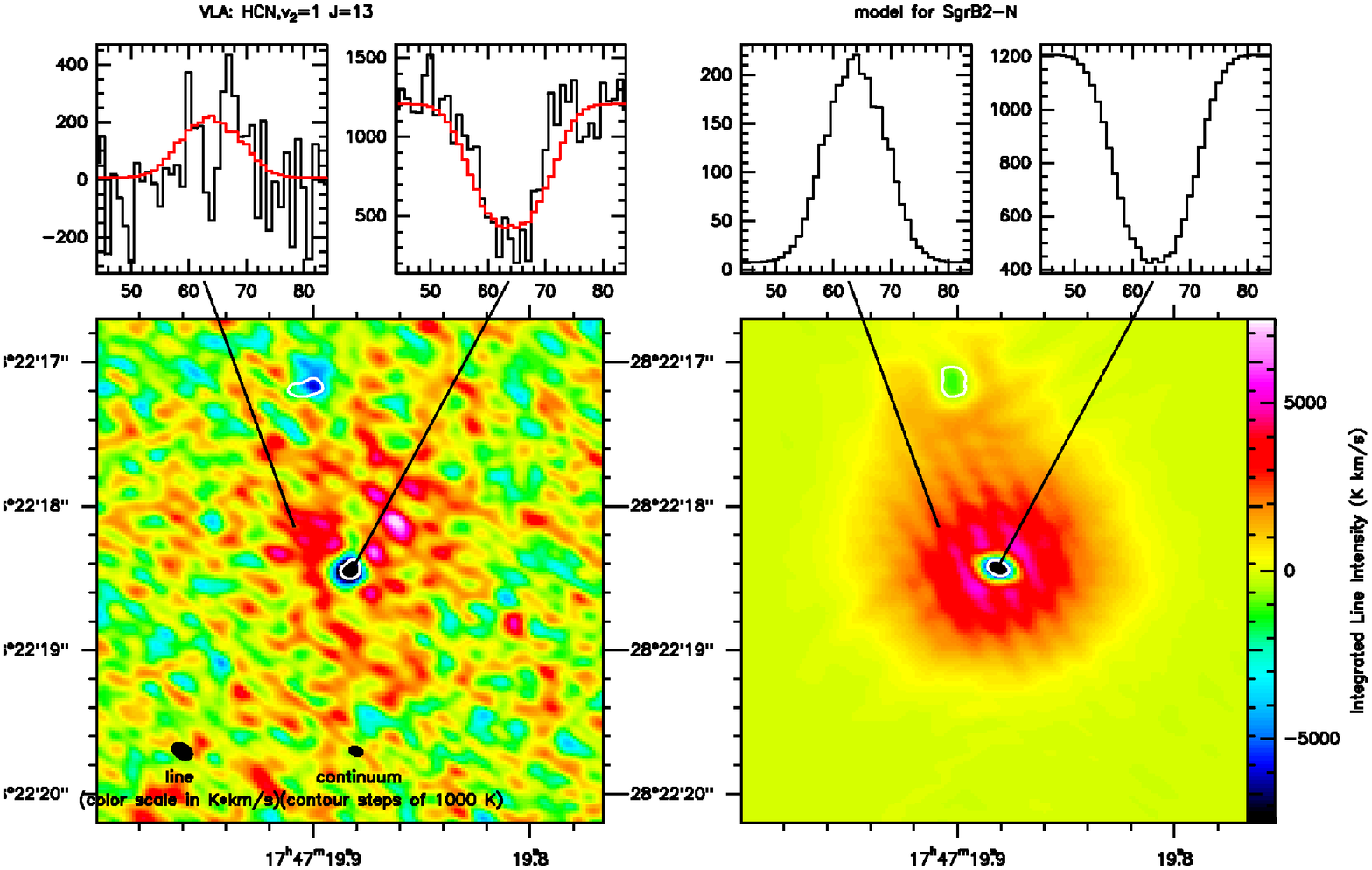}  
  \caption{Model for SgrB2-N (right) compared to the data (left). The model spectra are also overlaid in red. The white contours denote the continuum in steps of 1000 K.}
  \label{fig:b2n_mod}
\end{figure*}

\paragraph{SgrB2-N} In the model for SgrB2-N, the dust is heated by K2 with $10^5$ L$_\odot$ and K3 with $8.3\times 10^4$ L$_\odot$. The dust density follows a Gaussian centered at K2 with $7\times 10^7$ H$_2$ cm$^{-3}$ at the half-power radius of 10$^4$ AU. An additional core (Gaussian with $5\times 10^8$ H$_2$ cm$^{-3}$ at the half-power radius of 1000 AU) 3000 AU in front of K2 provides the strong absorption. The intrinsic line width is 10 km~s$^{-1}$. Figure~\ref{fig:b2n_mod} shows the comparison to the observations.

\begin{figure*}
  \centering
  \includegraphics[angle=-90,width=0.8\textwidth]{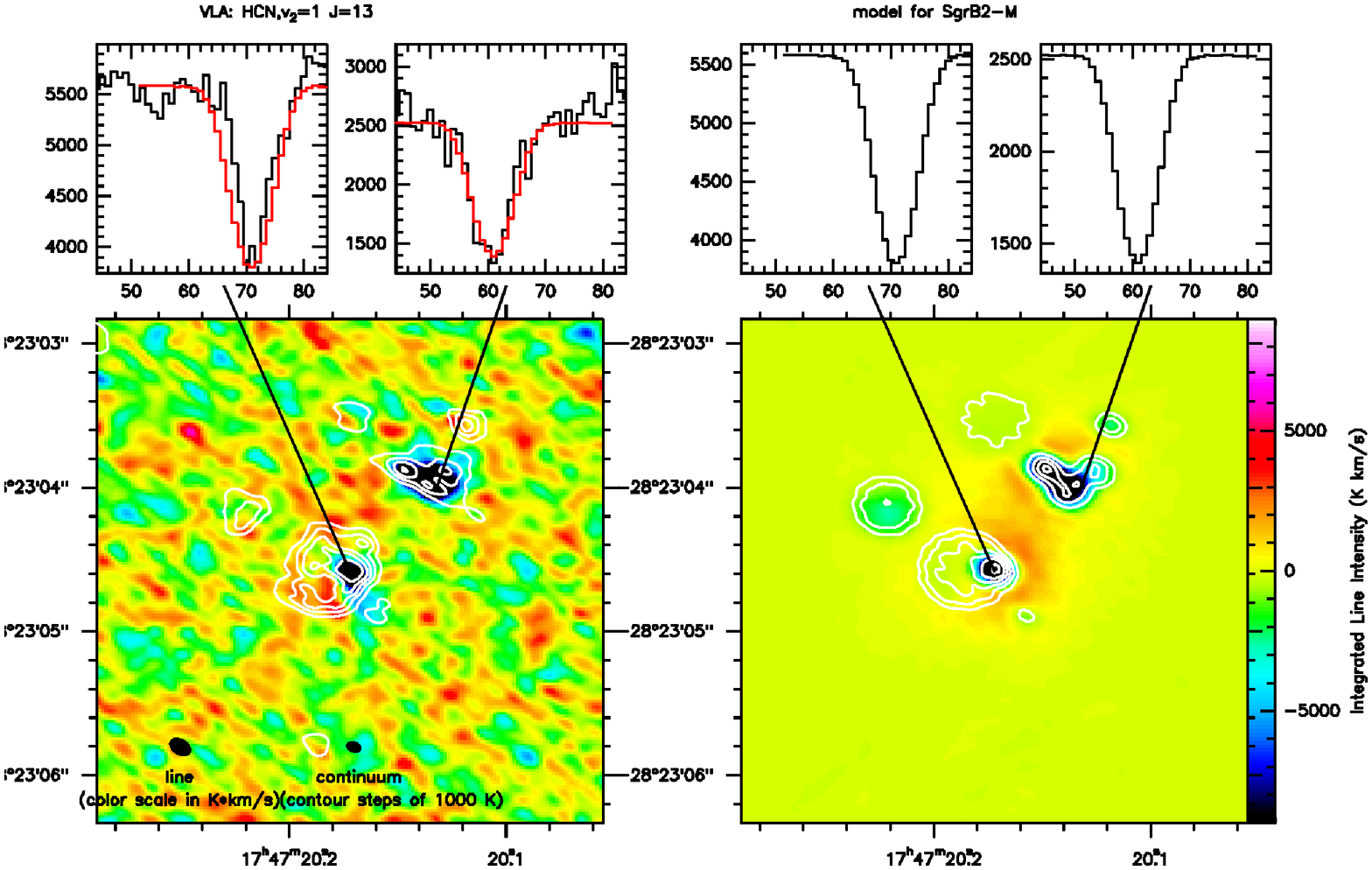}  
  \caption{Model for SgrB2-M (right) compared to the data (left). The model spectra are also overlaid in red. The white contours denote the continuum in steps of 1000 K.}
  \label{fig:b2m_mod}
\end{figure*}

\paragraph{SgrB2-M} In this model, the dust density follows two radial power laws with index 1.5, centered at F1f and F3. The first one starts with  $5.5\times 10^8$ H$_2$ cm$^{-3}$ at a radius of 660 AU, the latter one with $5\times 10^8$ H$_2$ cm$^{-3}$ at 750 AU. Inside this latter radius, there is an H{\sc ii} region that contains a star of $8\times 10^4$ L$_\odot$. Additionally, a larger  H{\sc ii} region is located 5000 AU closer to us with a star of $2.5\times 10^5$ L$_\odot$ - F3 is thus split into these two different H{\sc ii} regions. F1e with $2.5\times 10^4$ L$_\odot$ is located 2000 AU farther away from us than the other H{\sc ii} regions to facilitate absorption (Fig.~\ref{fig:coldens}). Further heating is provided by F1c, F1f, F2, F4 (each $5.4\times 10^4$ L$_\odot$), F1a ($3.8\times 10^4$ L$_\odot$), and F3a ($2.5\times 10^4$ L$_\odot$). The intrinsic line width is 6.7 km~s$^{-1}$. A comparison to the observations is shown in Fig.~\ref{fig:b2m_mod}.

These models are as simple as possible to approach a good fit to the observations, but are not unique. The model maps shown in Figs.~\ref{fig:g10_mod} to \ref{fig:b2m_mod} would appear more similar to the data if corresponding noise was added. This noise also makes it difficult to constrain more complicated models, since it is not clear how much of the clumpy substructure is real (which is on the order of only 2 or 3 times the rms noise)

%steps: ionized gas, heating, density distribution, abundance, compare, iterate to heating

%fig.: compmap, log. cuts in density and temperature

%spherical based on APEX (g10 sd)
%same density with 40 stars (lf2)
%higher density and bubble with 3 stars (pl2)

\section{Discussion}\label{sec:discussion}

In this section, we discuss implications from the observational results and the modeling efforts.

\subsection{Optical depth}

As explained in Sect.~\ref{sec:results}, lower limits on optical depth and HCN column density can be derived from the absorption lines (Table~\ref{tab:abslines}). Outstanding is K2 in SgrB2-N, where the line is optically thick. The column densities of hot HCN toward the H{\sc ii} regions where it is detected are at least a few times $10^{19}$ cm$^{-2}$, which translates into H$_2$ column densities of 10$^{24}$ to $10^{25}$ cm$^{-2}$ even for a high HCN fractional abundance of $10^{-5}$. This represents only the molecular gas at temperatures above roughly 300 K. In the models, the mass above 300 K is 350 M$_\odot$ in G10.47+0.03, 500  M$_\odot$ in SgrB2-N, and 200 M$_\odot$ in SgrB2-M.

Such large quantities of hot molecular gas are necessary to explain
the observations, and challenge theories of massive star
formation. Current hydrodynamical simulations produce far too little
hot molecular gas \citep[e.g.][with a total mass of 1000
  M$_\odot$]{Peters10}, both absolute and relative to the total mass,
since the highly inhomogeneous structure (filamentary disks) does not
lead to diffusion of the heating radiation. This is probably due to
the lower initial mass and hence lower column densities and dust optical depths of
these models, and it is to be hoped that in the future simulations of
cores as massive as the ones presented in the current paper will
reproduce our results. Also a high abundance of HCN (on the order of $10^{-5}$) in the dense, warm gas is needed, since the dust densities of our models are consistent with dust continuum data from the SMA and a lower abundance would mean an even higher mass of the hot gas. We expect future chemical models to compute such a high HCN abundance, probably through high-temperature gas-phase reactions.

Although HCN thermalizes to the ambient dust temperature, the population difference between the two $\ell$-type levels is very small, making them sensitive to the exact excitation conditions. The line optical depth is proportional to the population difference,
\begin{equation}\label{eq:popdiff}
\tau \propto N_{\rm l} - N_{\rm u}   \propto e^{-\frac{1411\ {\rm K}}{T_{\rm l}}} - e^{-\frac{1413\ {\rm K}}{T_{\rm u}}},  
\end{equation}
where $T_{\rm l}$ and $T_{\rm u}$ are the excitation temperatures for the lower and upper state, respectively. A difference of 0.1\% between $T_{\rm l}$ and $T_{\rm u}$ causes $\tau$ to change by more than 70\%; inversion occurs if $T_{\rm u}$ is 0.14\% larger than $T_{\rm l}$.  The two levels are independently coupled to the ground vibrational state through the 14$\mu$m transitions, which determine the excitation temperatures. If these are different, the analysis would be far more difficult, as one would have to know the exact 14$\mu$m  radiation field,  where lines could slightly affect the optical depths. However, there is no such mechanism known, and an extreme sub-thermal population difference, which would greatly reduce the required absorption column densities, is highly unlikely because of the agreement between absorption and emission, between $J$=13 and other direct $\ell$-type lines (Fig.~\ref{fig:rd}), and between the derived column densities and dust emission data. Therefore, we consider LTE a good assumption for this transition.

If we assume a similar optical depth for the emission lines, say 0.3, and a temperature of 400 K, then the line intensity should be 100 K, which is just the noise level in 1 km~s$^{-1}$ channels and 0.13$''$ beams. Owing to the strength of the free-free radiation, we are thus more sensitive to absorption than emission lines. Averaging over a larger region (Fig.~\ref{fig:spectra} and Table~\ref{tab:emlines}) lowers both noise and peak intensities. The latter reach up to 300 K in G10.47+0.03, suggesting that optical depth and temperature are higher than the above mentioned assumptions. Surprising is the detection of the line toward G10.47+0.03 A, since no NH$_3$(4,4) satellites were detected by \citet{Cesaroni10}.

In the model for SgrB2-M, we have placed the H{\sc ii} regions F3 and F1e in front of and behind the bulk of the molecular gas, respectively. This prevents absorption toward most of F3 and leads to stronger absorption toward F1e. The exact offsets in z are not well constrained, however. To obtain the very strong absorption toward K2 in SgrB2-N, we put a very dense core just in front of the H{\sc ii} region. This is of course not very satisfying as it is unlikely that such a core is exactly along the line-of-sight. It was however not possible to reproduce both emission and absorption with a more symmetric distribution.

\subsection{Heating}

Heating up large masses of molecular gas requires deeply embedded massive (proto)stars with high luminosities. Their radiation is either originally in the infrared \citep{Hosokawa09}, not producing H{\sc ii} regions, or is quickly processed to the infrared by dust absorption of the UV radiation. Due to high column densities in all directions, the dust is optically thick even in the infrared and this radiation cannot escape, but diffuses outwards by multiple absorption/emission events until the dust is optically thin to its own radiation. This diffusion (or radiative trapping) leads to much higher temperatures in the inner part; e.g. at 2000 AU from the HCH{\sc ii} region B1, diffusion raises the temperature from 150 K in a low-density model, where diffusion is not effective, to 450 K  in the high-density model for G10.47+0.03. This mechanism therefore seems indispensable to account for large masses of hot gas. Additionally, by absorbing stellar UV radiation the dust protects the molecules from dissociation. In general, heating is more efficient if the heating source is inside a density condensation \citep{Kaufman98} than if it is external, as could be the case for G34.26+0.15 \citep{Watt99, Mookerjea07}.
% a steeper temperature gradient (where there is also a significant density gradient) and hence

In our models, the diffusion mechanism is accomplished by dense spherical clumps around heating sources.  It is however not clear how many heating sources there are: The stars in the H{\sc ii} regions are sufficient to reproduce the observed line strength, as the presented models demonstrate. But a good fit can also be achieved when lower-luminosity stars are added to the models according to the Initial Luminosity Function, without changing the stars in the H{\sc ii} regions. So on the basis of these data it seems not possible to distinguish between the scenarios of heating only by the stars ionizing the H{\sc ii} regions and heating by H{\sc ii} regions plus embedded sources which do not emit in the radio regime.  We may speculate that there are much more stars present than detected from the free-free radiation, since the observed sources are clusters in formation. Also, there could be more extended (faint) line emission (and the absorption coming from a larger column) than in the models, which would point to additional heating sources. Furthermore, the simple models predict strong self-absorption in the (optically thick) submm rotational transitions within the $v_2$=1 state of HCN and H$^{13}$CN, which is however not observed with the SMA, presumably due to clumpiness (Rolffs et al. 2011, in prep.).

While not excluding the possibility of additional heating sources, we could show that heating by the stars in the hypercompact H{\sc ii} regions is sufficient to reproduce the observations. Hence, our modeling results suggest that it is unlikely that the main heating is due to embedded objects without ionized gas. Given that the sources are dense condensations deeply embedded in giant molecular clouds and have high column densities, diffusion of radiation, which is necessary to increase the heating, is reasonable even for more inhomogeneous and asymmetric structures than modeled. 

% clumpy -> no strong abs? no, just scatter in N

%Only heating by stars in the H{\sc ii} regions that are visible in the continuum map is taken into account. This is sufficient to reproduce the observed line strength. Another possibility would be additional heating sources. We tried to add lower-luminosity stars to the models according to the Initial Luminosity Function. A result of this exercise is that a fit can be achieved as well, so on the basis of these data it seems not possible to distinguish between the scenarios of heating only by the H{\sc ii} regions and heating by H{\sc ii} regions plus embedded sources which do not emit in the radio. 

%In this picture, the distribution of optical depth follows the density distribution; an inhomogeneous line emission would reflect a clumpy structure. However, the density in the models is homogeneous on small scales, so the line emission is as well.

%However, it is clear that the observed sources are clusters in formation, so there must be much more stars than visible from  the strong free-free emission. Also, submm data from the SMA contradict the simple models presented in Section~\ref{sec:modeling}, since they predict strong self-absorption in the optically thick rotational transitions within the $v_2$=1 state of HCN, which is not observed (Rolffs et al., in prep.). In SgrB2-N, the line emission could be more extended (and the absorption coming from a larger column), which would point to additional heating sources. 

%\subsection{Density Distribution}

\subsection{Velocity field}

Many systematic motions are hidden in the line width of around 10 km~s$^{-1}$, which contains both turbulent and macroscopic motions. These components (e.g. the presence of rotation, infall, or outflows) can not be disentangled.  For all emission lines, the signal-to-noise ratio is too poor to gain accurate information on the velocity structure.

%free-fall velocity at 1000 AU with 20 M$_\odot$ would be ${\rm v}=- \sqrt{2 G M_{\rm in}/r} = - 1.6$ km/s
%The linewidth can at least give you an upper limit

%why not simply a second velocity component?

\paragraph{G10.47+0.03} The most striking feature in this source is an additional blue-shifted absorption component in both B1 and B2 at a velocity of around 60 km~s$^{-1}$, compared to the source velocity of 68 km~s$^{-1}$ (Fig~\ref{fig:spectra}). %The blue-shifted absorption also shows up in the NH$_3$(4,4) line \citep{Cesaroni98, Cesaroni10} and in  ground-state transitions (Rolffs et al. 2011, in prep.). 
The NH$_3$(4,4) absorption \citep{Cesaroni98, Cesaroni10} is at 53 km~s$^{-1}$ toward B1 and at 61 km~s$^{-1}$ toward B2, with much lower optical depth than the emission component. An explanation could be that the colder gas, probed by NH$_3$(4,4) with its level energy of 200 K, is mainly in an outflow cone, while the hot gas probed by HCN,$v_2$=1,$J$=13 is mainly at the systemic velocity close to the HCH{\sc ii} regions and only partly expanding. With the SMA, we see broad blue-shifted absorption in several lines toward the dust continuum peak between B1 and B2 (Rolffs et al., in prep.). This indicates expansion motions involving many different molecular excitation conditions, while at the same time large-scale infall is present \citep{Rolffs11apex}. Given the high excitation and the different velocities, an unrelated foreground velocity component is unlikely. 
The region west of B1 appears to move at a velocity of 67 km~s$^{-1}$, while south of B1 the velocity is 69 km~s$^{-1}$.

\paragraph{SgrB2-N} While the strong absorption line toward K2 has a line width of 10 km~s$^{-1}$ at the source velocity of 64 km~s$^{-1}$, the emission lines north of K2 have a higher line width and higher velocity, possibly indicating an additional component at $\sim$70 km~s$^{-1}$.

\paragraph{SgrB2-M} Three velocity components can be seen from the absorption lines (Fig.~\ref{fig:b2m_pv}), one at around 50 km~s$^{-1}$ in F1f, one at 60 km~s$^{-1}$ in F1c, F1e, and F1f, and one at around 70 km~s$^{-1}$ in F3. Relative to the ambient velocity of 60 or 64 km/s, which itself is not well determined due to different components, the F1f line shows blue-shifted absorption, indicating an expansion component, and the F3 line is red-shifted. This could mean infall or a motion of the whole F3 core. The F3 line shows a northeast-southwest velocity gradient from 69 to 73 km~s$^{-1}$ over 0.3$''$ (2300 AU). If that is interpreted as Keplerian rotation, and if it is assumed that the whole orbit is seen and is edge-on, the lower limit on the central mass would be 5 M$_\odot$. The large-scale velocity field shows expansion in the inner parts and infall in the outer parts \citep{Rolffs10}.

%\subsection{}

\section{Conclusions}\label{sec:concl}

With the VLA, we have obtained maps of a high-excitation (1400 K) HCN line at high angular resolution (0.1$''$) toward three massive star-forming regions. We see both absorption toward small and dense H{\sc ii} regions and emission. The infered optical depths require on the order of $10^{24}$ H$_2$ cm$^{-2}$ and 100 solar masses of hot ($>$300 K) molecular gas within a region of less than 0.1 pc. We note that the column densities of hot HCN derived from the absorption lines are only lower limits, and can be substantially higher especially due to beam dilution. Heating is provided by the stars powering hypercompact H{\sc ii} regions, possibly complemented by heating sources invisible in the radio domain. Diffusion of radiation in an optically thick dusty environment is likely to increase the temperatures. The line velocities reveal expansion motions in G10.47+0.03 and components at 50, 60, and 70 km~s$^{-1}$ in SgrB2-M, the latter with a velocity gradient. Our three-dimensional radiative transfer modeling with RADMC-3D demonstrate the great power and potential of our approach to constrain the source structure of high-mass star-forming regions. HCN direct $\ell$-type lines are a unique tool, whose future observation will also be possible with ALMA and the SKA.

%\section{Summary}

\begin{acknowledgements}
We thank the VLA staff for having carried out the observations. ST acknowlegdes support by the Deutsche Forschungsgemeinschaft through grant TH 1301/3-1.
%We thank Arnaud Belloche for the 30m data of HCN direct $\ell$-type transitions.
\end{acknowledgements}

\bibliographystyle{aa}
\bibliography{16544ref}

\end{document}